\documentclass[aps,prb,reprint,groupedaddress]{revtex4-2}

\usepackage{graphicx}

\begin{document}



\title{Dissipative conductivity of a dirty superconductor with Dynes subgap states under a dc bias current up to the depairing current density}


\author{Takayuki Kubo}
\email[]{kubotaka@post.kek.jp}
\affiliation{High Energy Accelerator Research Organization (KEK), Tsukuba, Ibaraki 305-0801, Japan}
\affiliation{The Graduate University for Advanced Studies (Sokendai), Hayama, Kanagawa 240-0193, Japan}



\begin{abstract}
We study the dissipative conductivity $\sigma_1$ of a dirty superconductor with a finite Dynes parameter $\Gamma$ under a dc-biased weak time-dependent field. 
The Usadel equation for the current-carrying state is solved to calculate the pair potential, penetration depth, supercurrent density, and quasiparticle spectrum. 
It is shown that, while the depairing current density $j_d$ for $\Gamma=0$ is coincident with the Kupriyanov-Lukichev theory, 
a finite $\Gamma$ decreases the superfluid density, resulting in a reduction of $j_d$. 
The broadening of the peaks of the quasiparticle density of states induced by a combination of a finite $\Gamma$ and a dc bias can reduce $\sigma_1$ below that for the ideal dirty BCS superconductor with $\Gamma=0$, 
while subgap states at Fermi level proportional to $\Gamma$ results in a residual conductivity at $T\to 0$. 
We find the optimum combination of $\Gamma$ and the dc bias to minimize $\sigma_1$ by scanning all $\Gamma$ and all currents up to $j_d$. 
By using the results, it is possible to improve $j_d$ and reduce electromagnetic dissipation in various superconducting quantum devices.
\end{abstract}

\maketitle


\section{Introduction}
Electromagnetic properties of superconductors have been actively studied in many fields of fundamental and applied physics, 
including applications to superconducting radio-frequency (SRF) cavities for particle accelerators~\cite{2017_Padamsee, 2017_Gurevich_SUST}, microresonators for kinetic inductance detectors~\cite{2012_Zmuidzinas} and quantum computations~\cite{2013_Devoret}, and single-photon detectors~\cite{2015_Engel}, etc. 
One of the striking features of superconductivity in applied perspectives is the ultra law dissipation in $s$-wave superconductors at temperatures $T$ well below the critical temperature $T_c$ and photon frequencies $\hbar \omega_{\gamma}$ smaller than the superconducting gap $\Delta$. 
For instance, modern niobium SRF cavities exhibit~\cite{2017_Padamsee, 2014_Romanenko, 2017_Romanenko} surface resistance $R_s < 10\,{\rm n\Omega}$ or quality factors $> 10^{10}$ at $T \lesssim 2\,{\rm K}$ and $\omega_{\gamma}/2\pi \sim 1\,{\rm GHz}$ under weak and strong rf currents close to the depairing current density $j_d \sim H_c/\lambda$. 
Here $H_c$ is the thermodynamic critical field and $\lambda$ the penetration depth.

A quality factor of the superconducting resonator is proportional to $1/R_s \propto 1/\sigma_1$. 
Here the dissipative conductivity $\sigma_1$ is the real part of complex conductivity, 
which is sensitive to the details of the quasiparticle spectrum. 
The quasiparticle density of states (DOS) of the ideal BCS superconductor in the zero-current state is given by
$N(\epsilon) = N_0 \epsilon/\sqrt{\epsilon^2 -\Delta^2}$, 
where $N_0$ is the density of states at the Fermi level in the normal state.  
In this case, $\sigma_1$ is calculated from the Mattis-Bardeen (MB) formula~\cite{1958_MB}
$\sigma_1/\sigma_n = (2\Delta/k_B T) \ln (4e^{-\gamma_E} k_B T / \hbar \omega_{\gamma}) \exp(-\Delta/k_B T)$  
for a dirty superconductor at $\hbar \omega_{\gamma} \ll k_B T \ll \Delta$. 
Here $\gamma_E=0.577$ is the Euler constant. 
However, as revealed in many tunneling experiments~\cite{2003_Zasa}, 
quasiparticle DOS has a finite density of subgap states at $|\epsilon | < \Delta$ and the DOS peaks at $\epsilon = \Delta$ are smeared out. 
Such DOS has been described by the Dynes formula~\cite{1978_Dynes, 1984_Dynes}, 
$N(\epsilon) = N_0 {\rm Re} [(\epsilon + i\Gamma)/\sqrt{(\epsilon + i\Gamma)^2 -\Delta^2}]$, 
where $\Gamma$ is a phenomenological parameter to describe the broadening of the DOS peaks. 
It is also well-known that the pair-breaking mechanisms such as the Meissner currents~\cite{1969_Maki_Parks, 1964_Maki_current,1965_Fulde_current, 2003_Anthore}, magnetic impurities~\cite{1969_Maki_Parks, 1966_Fulde_Maki_mag}, a proximity-coupled normal layer~\cite{1996_Belzig, 1999_Belzig}, etc. also broaden the DOS peaks. 
Unfortunately, these realistic cases are outside the scope of the simple MB formula with the ideal BCS DOS.

The effects of the rf field on $\sigma_1$ are studied based on the more general formula derived using the Keldysh technique of the nonequilibrium Green's function~\cite{2014_Gurevich, 2016_Semenov}. 
It was shown~\cite{2014_Gurevich} that the broadening of the DOS peaks due to the strong rf field $H$ with $\hbar \omega_{\gamma} \ll k_B T$ can reduce $R_s$ and results in a pronounced minimum in $R_s(H)$. 
We can qualitatively understand this result by looking back at the MB formula. 
The logarithmic divergence at $\omega_{\gamma} \to 0$ in the MB formula comes from the sharp DOS peaks at $\epsilon=\Delta$ in the BCS DOS. 
When the current-induced broadening of the DOS peaks is given by $\delta \epsilon > \hbar \omega_{\gamma}$, 
the denominator in the logarithmic factor is replaced with $\delta \epsilon$ and the divergence at $\omega_{\gamma} \to 0$ disappears. 
As $\delta \epsilon$ increases, $\sigma_1$ is logarithmically decreased, 
consistent with the experiment~\cite{2014_Ciovati_Dhakal_Gurevich}. 
On the other hand, the reduction of the spectrum gap $\delta \epsilon$ increases $\sigma_1$. 
The interplay of the broadening of the DOS peaks and the reduction of the spectrum gap determines the minimum of $\sigma_1 (H)$.

Pair-breaking effects due to realistic materials features including magnetic impurities, Dynes $\Gamma$ parameters, and a proximity-coupled normal layer at the surface can also reduce $R_s$ via the broadening of the DOS peaks~\cite{2017_Gurevich_Kubo}. 
For instance, sparse magnetic impurities can reduce $R_s$ by $\sim 50\%$ for the weak rf field. 
More recently, it was shown~\cite{2019_Kubo_Gurevich} that a combination of such pair-breaking effects in materials and the pair-breaking current can shift the minimum in nonlinear $R_s (H)$,  
consistent with the experimental observations that the nonlinear behavior of $R_s$ is sensitive to materials treatments~\cite{2012_Antoine, 2013_Grassellino, 2013_Dhakal, 2017_Grassellino, 2017_Maniscalco, 2018_Dhakal,2019_Wenskat, 2019_Umemori}.

These studies suggest the engineering of the DOS using various pair-breaking mechanisms can minimize dissipation in superconducting devices. 
The dc bias current or field is a convenient control knob for tuning the quasiparticle spectrum~\cite{2014_Gurevich}. 
From applied perspectives, studying $\sigma_1$ under the dc current $j_s$ superposed on the weak time-dependent field and reveal the optimum $j_s$ to minimize dissipation would attract attention in superconducting device communities. 
From fundamental perspectives, 
this system offers a stage for direct observations of the effects of the broadening of the DOS peaks on $\sigma_1$~\cite{2014_Gurevich, 2017_Makita, 2019_Maniscalco}. 
In measurements under the strong rf current, on the other hand, these effects are mixed up with the slow dynamics of nonequilibrium quasiparticles that control the distribution function~\cite{1981_WattsTobin, Kopnin}. 
In this paper, we consider a superconductor with Dynes subgap states. 
The Dynes $\Gamma$ has not been derived from a microscopic theory, 
yet we can incorporate a finite $\Gamma$ into the quasiclassical theory of the BCS model~\cite{2017_Gurevich_SUST, 2017_Gurevich_Kubo, 2019_Kubo_Gurevich}. 
We study the effect of a combination of $\Gamma$ and the dc bias for all $\Gamma$ and all currents up to the deparing current density $j_d$. 
To do so, we need to calculate $j_d$ for $\Gamma>0$. 
Although $j_d$ of dirty-limit superconductors for $\Gamma=0$ was calculated many years ago~\cite{1980_Kupriyanov}, 
that for $\Gamma>0$ is still unknown. 
The value of $j_d (\Gamma, T)$ is related to the maximum accelerating field that SRF cavities can achieve with the bulk SRF~\cite{2008_Catelani, 2012_Lin_Gurevich, 2015_Kubo_PTEP, 2017_Kubo_SUST, 2017_Liarte_SUST, 2019_Sauls, 2015_Posen_PRL} and the thin-film SRF technologies~\cite{2006_Gurevich, 2014_Kubo, 2015_Gurevich, 2017_Kubo_SUST, 2016_Tan}, 
and also related to the threshold current of superconducting nanowire single-photon detectors~\cite{2015_Engel}.

The paper is organized as follows. 
In Section II, we briefly review the quasiclassical theory for a dirty superconductor. 
We express various physical quantities with the Matsubara Green's functions and the retarded Green's functions. 
In Sec. III, we evaluate the effects of $\Gamma$ on $T_c$, $\Delta$, superfluid density $n_s$, $\lambda$, $N(\epsilon)$, and $\sigma_1$ in the zero-current state. 
In Sec. IV, we calculate $\Delta$, $n_s$, and $\lambda$ in the current-carrying state and express the supercurrent density $j_s$ as a function of superfluid momentum. 
The maximum value of $j_s$ is the depairing current density $j_d(\Gamma,T)$. 
Then we investigate the effects of $\Gamma$ on $j_d$ for all $T$. 
By using these results, we evaluate the effects of $\Gamma$ and $j_s$ ($\le j_d$) on the DOS. 
Then we consider the case that the dc bias $j_s$ is superposed on the weak time-dependent current with the frequency $\omega_{\gamma}$ and calculate  $\sigma_1 (j_s, \Gamma, T, \omega_{\gamma})$. 
In Sec. V, we discuss the implications of our results.

\section{Theory}

We use the well-established quasiclassical formalism for the dirty limit, the Usadel equation~\cite{1970_Usadel, Kopnin}. 
Consider a dirty superconductor in which the current varies slowly over the coherence length. 
Then the local values of the normal and anomalous quasiclassical Matsubara Green's functions $G=\cos\theta$ and $F=\sin\theta$ obey 
\begin{eqnarray}
s \sin\theta \cos \theta + (\hbar \omega_n + \Gamma) \sin \theta - \Delta \cos\theta = 0 . \label{thermodynamic_Usadel}
\end{eqnarray}
Here $s =  (q/q_{\xi})^2 \Delta_0$, 
$\Delta_0 = \Delta(s, \Gamma, T)|_{s=\Gamma=T=0}$ the BCS pair potential at $T=0$, 
$\hbar q = 2 m v_s$ the superfluid momentum, 
$v_s$ the superfluid velocity, $m$ the electron mass, 
$q_{\xi} = \sqrt{2\Delta_0/\hbar D}$ the inverse of the coherence length, 
$D=\sigma_n/2 e^2 N_0$ the electron diffusivity, 
and $\hbar \omega_n = 2\pi k_B T(n + 1/2)$ the Matsubara frequency. 
The pair potential $\Delta$ satisfies the self-consistency equation
\begin{eqnarray}
\ln \frac{T_{c0}}{T} = 2 \pi k_B T \sum_{\omega_n >0} \biggl( \frac{1}{\hbar \omega_n} - \frac{\sin\theta}{\Delta} \biggr) ,
\label{self-consistency}
\end{eqnarray}
where $k_B T_{c0}= \Delta_0 \exp(\gamma_E)/\pi  \simeq \Delta_0/1.76$ is the BCS critical temperature, 
and $\gamma_E=0.577$ is the Euler constant. 
The superfluid density $n_s$, penetration depth $\lambda$ and supercurrent density $j_s$ are given by
\begin{eqnarray}
&&\frac{n_s(s, \Gamma, T)}{n_{s0}} = \frac{4k_B T}{\Delta_0} \sum_{\omega_n > 0} \sin ^2 \theta 
, \label{superfluid_density} \\
&&\lambda^{-2}(s, \Gamma, T) 
= \frac{\mu_0 e^2 n_s}{m}  
= \frac{n_s(s, \Gamma, T)}{n_{s0}} \lambda_0^{-2} , \label{lambda} \\
&&j_s (s, \Gamma, T) 
= -en_s v_s 
= \frac{n_s(s, \Gamma, T) }{n_{s0}} \sqrt{\frac{\pi s}{\Delta_0}}  j_{s0} , \label{supercurrent} 
\end{eqnarray}
Here $n_{s0} = n_s(0, 0, 0) = 2\pi m N_0 D \Delta_{0}/\hbar$ is the BCS superfluid density at $T=0$, 
$\lambda_0 = \lambda(0,0,0) = \sqrt{\hbar/\pi \mu_0 \Delta_0 \sigma_n}$ the BCS penetration depth at $T=0$, 
$j_{s0} = H_{c0}/\lambda_0 = -\sqrt{\pi} e N_0 D \Delta_0 Q_{\xi}$, 
and $H_{c0}=\sqrt{N_0/\mu_0} \Delta_0$ the BCS thermodynamic critical field at $T=0$.

To calculate $N(\epsilon)$ and $\sigma_1$, 
we need the retarded normal and anomalous Green's functions $G^R=\cosh(u+iv)$ and $F^R=\sinh(u+iv)$, 
where $u$ and $v$ satisfy the real-frequency Usadel equation 
\begin{eqnarray}
&&is \sinh(u+iv) \cosh (u+iv) \nonumber \\
&&+ (\epsilon + i\Gamma) \sinh (u+iv) - \Delta \cosh(u+iv) = 0 . \label{real_freq_Usadel}
\end{eqnarray}
The quasiparticle DOS is given by
\begin{eqnarray}
\frac{N(\epsilon)}{N_0} = {\rm Re} G^R = \cosh u \cos v . \label{DOS}
\end{eqnarray}
and $\sigma_1(s, \Gamma, T, \omega_{\gamma})$ is given by~\cite{2014_Gurevich}
\begin{eqnarray}
\frac{\sigma_1}{\sigma_n} = \frac{1}{\hbar \omega_{\gamma}} \int_{-\infty}^{\infty}\!\!\! d\epsilon [f(\epsilon)-f(\epsilon + \hbar \omega_{\gamma})] M(\epsilon, \Gamma, \omega_{\gamma}, s) , \label{sigma_1}
\end{eqnarray}
where $f$ is the quasiparticle distribution function and $M$ the spectral function 
\begin{eqnarray}
M = {\rm Re} G^R(\epsilon) {\rm Re} G^R(\epsilon + \hbar \omega_{\gamma} ) + {\rm Re} F^R(\epsilon) {\rm Re} F^R(\epsilon + \hbar \omega_{\gamma} ) . \nonumber \\
\end{eqnarray}
In general, $f$ is determined by nonequilibrium dynamics of quasiparticles~\cite{1981_WattsTobin, Kopnin}. 
For the cases contributions from nonequilibrium quasiparticles are negligible, 
$f$ is given by the Fermi distribution $f=(\exp(\epsilon/k_B T)+1)^{-1}$, 
yielding the well-known formula~\cite{2012_Clem_Kogan, 1967_Nam}. 
In this work, we study $\sigma_1$ for the weak-field limit with and without the dc current, 
in which nonequilibrium dynamics of quasiparticles driven by the time-dependent current is negligible, 
so we use the Fermi distribution function. 
The imaginary part of the complex conductivity can be calculated from $\sigma_2 = 1/\mu_0 \omega_{\gamma}\lambda^{2}(s,\Gamma,T)$ for $\sigma_1 \ll \sigma_2$. 
Here $\lambda$ is given by Eq.~(\ref{lambda}).

In the following, we use $\Delta_{0}$ as a unit of energy and use dimensionless quantities 
$\tilde{s} = s/\Delta_{0}$, $\tilde{\omega}_n= \hbar \omega_n/\Delta_{0}$, 
$\tilde{\omega}_{\gamma}= \hbar \omega_{\gamma}/\Delta_{0}$, 
$\tilde{\Gamma} = \Gamma/\Delta_{0}$, $\tilde{\Delta}= \Delta/\Delta_{0}$, 
$\tilde{T}= k_B T/\Delta_{0}$, etc. 
For brevity, we omit all these tildes. 

\section{Zero-current state} \label{zero_current_state}

\begin{figure}[tb]
   \begin{center}
   \includegraphics[width=0.48\linewidth]{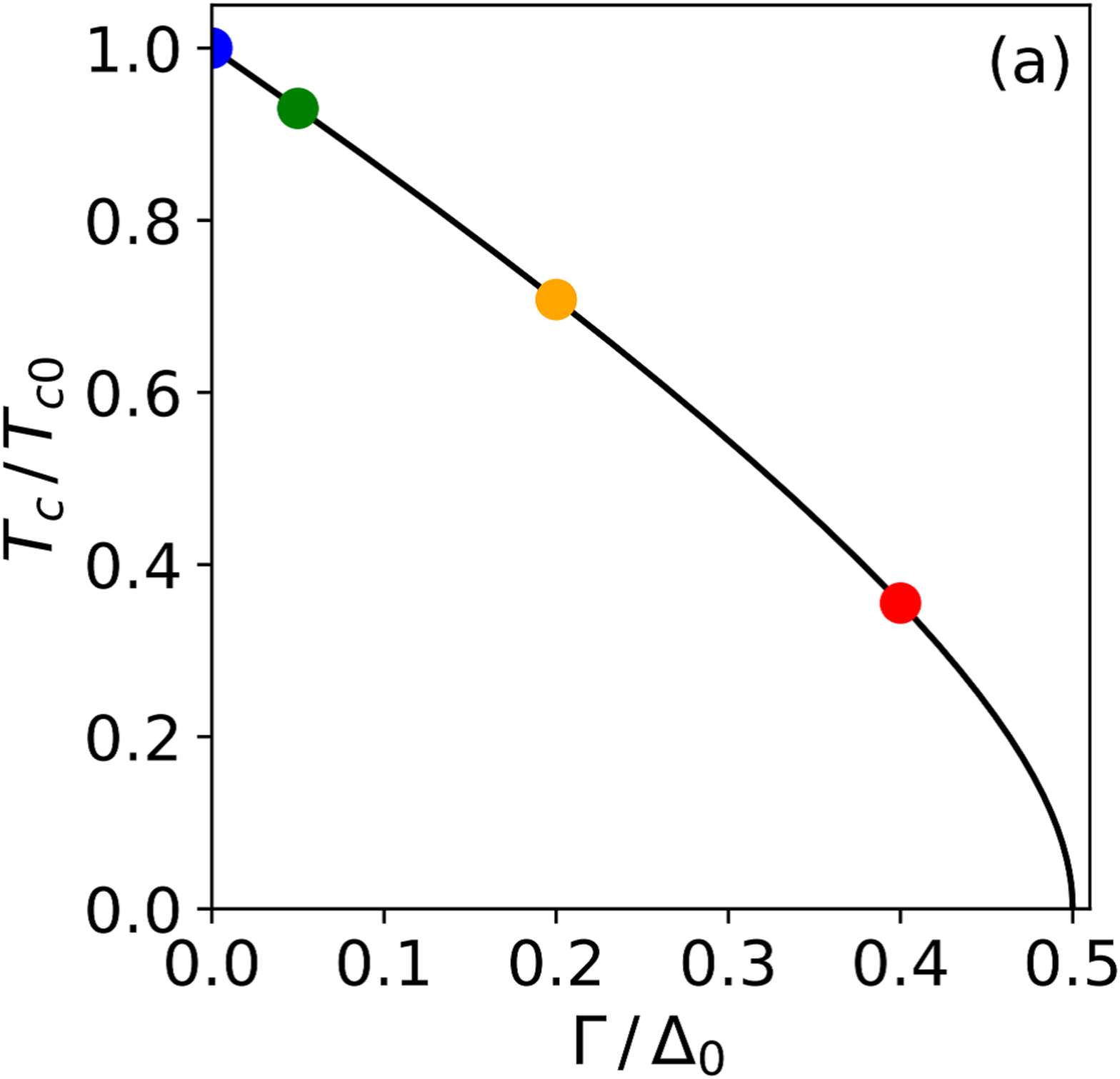}
   \includegraphics[width=0.48\linewidth]{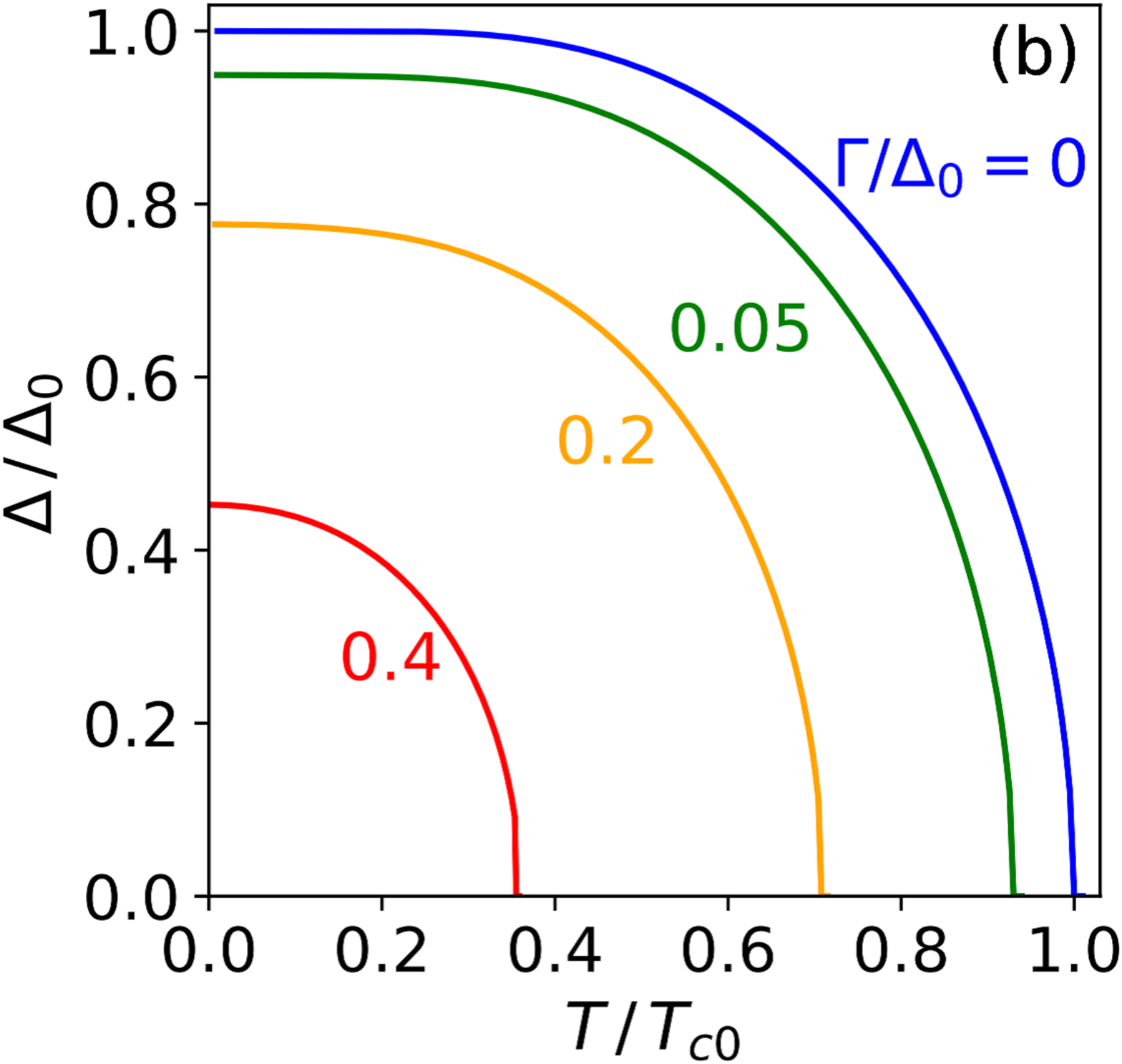}
   \includegraphics[width=0.48\linewidth]{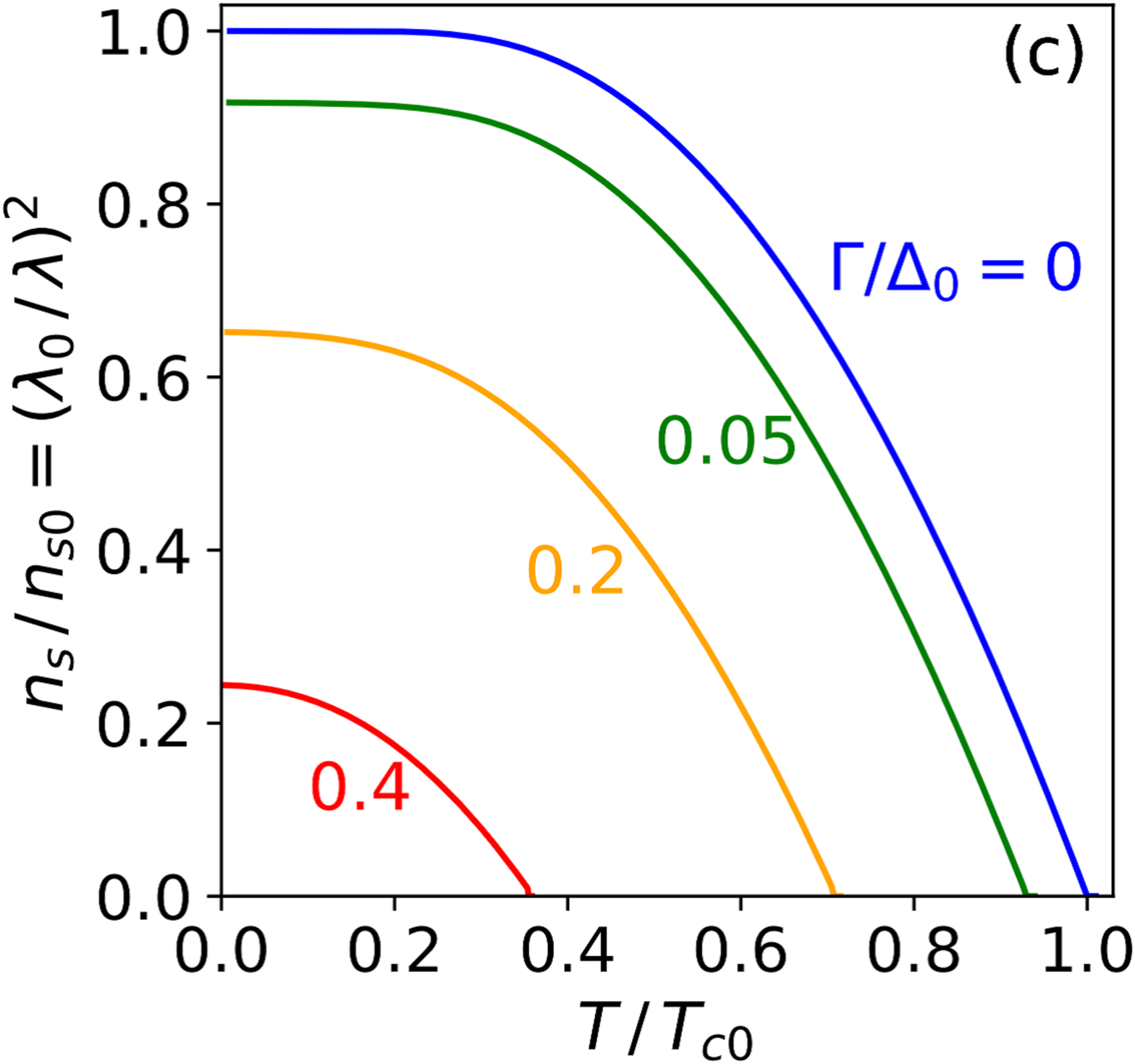}
   \includegraphics[width=0.48\linewidth]{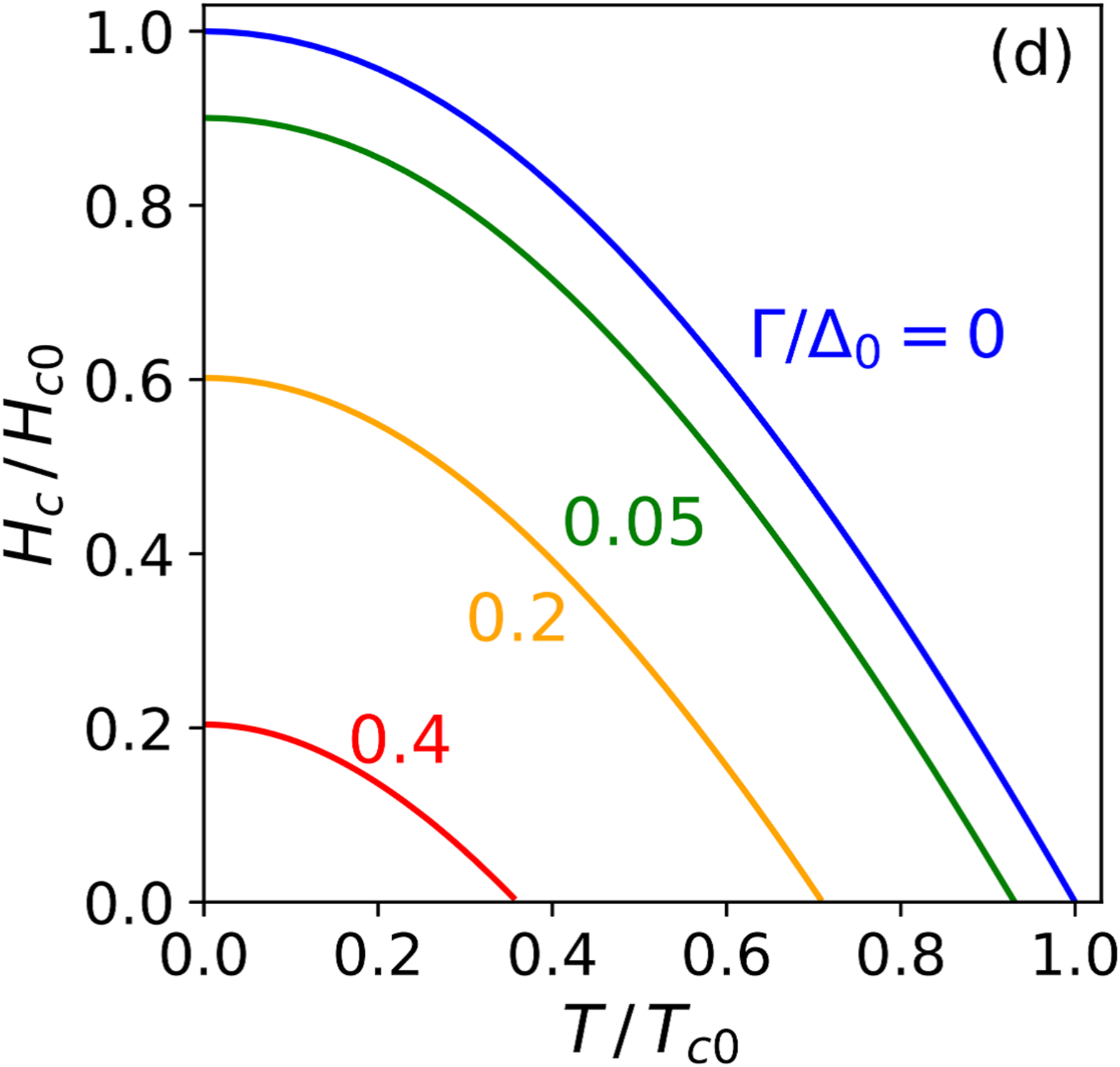}
   \end{center}\vspace{0 cm}
   \caption{
Thermodynamic properties in the zero-current state. 
(a) Critical temperature $T_c (\Gamma)$. 
For instance, $T_c/T_{c0}=1, 0.93, 0.71, 0.36$ for $\Gamma=0, 0.05, 0.2, 0.4$, respectively. 
(b) Pair potential $\Delta(0, \Gamma,T)$. 
(c) Superfluid density $n_s (0, \Gamma, T)$ and penetration depth $\lambda^{-2}(0, \Gamma, T)$. 
(d) Thermodynamic critical field $H_c(\Gamma, T)$.
   }\label{fig1}
\end{figure}

First consider the zero-current state ($s \propto q^2 \to 0$). 
Solving Eqs.~(\ref{thermodynamic_Usadel}) and (\ref{self-consistency}) for $(\theta, \Delta) \ll 1$, 
we obtain the equation for the critical temperature $T_c(\Gamma)$~\cite{2017_Gurevich_Kubo}
\begin{eqnarray}
\ln \frac{T_{c}}{T_{c0}} = \psi \biggl(\frac{1}{2} \biggr) - \psi\biggl( \frac{1}{2} + \frac{\Gamma}{2\pi T_{c}} \biggr) , \label{Tc} 
\end{eqnarray}
Here $\psi$ is the digamma function. 
Note here Eq.~(\ref{Tc}) has the same form as the well-known equation for the critical temperature of a superconductor with pair-breaking perturbations~\cite{1964_Maki_current, 1966_Fulde_Maki_mag}. 
Expanding the digamma function about $1/2$ yields a formula $T_c(\Gamma) = T_{c0} - \pi \Gamma/4$ for $\Gamma \ll 1$. 
The numerical solution of Eq.~(\ref{Tc}) gives $T_c$ for an arbitrary $\Gamma$. 
As shown in Figure~\ref{fig1} (a), 
$T_c$ monotonically decreases with $\Gamma$ and vanishes at $\Gamma = 1/2$.

The solution of Eq.~(\ref{thermodynamic_Usadel}) is given by $\sin\theta_{\Gamma} = \Delta/\sqrt{(\omega_n +\Gamma)^2+\Delta^2}$ where $\Delta$ satisfies Eq.~(\ref{self-consistency}). 
At $T \to 0$, the summation in Eq.~(\ref{self-consistency}) is replaced with integration, 
which yields $\Delta (0, \Gamma,T)|_{T\to 0}= \sqrt{1-2\Gamma}$ or $\simeq 1-\Gamma$ for $\Gamma \ll 1$. 
For an arbitrary $T$, we need to solve Eqs.~(\ref{thermodynamic_Usadel}) and (\ref{self-consistency}) numerically. 
Shown in Fig.~\ref{fig1} (b) is $\Delta (0, \Gamma, T)$ as functions of $T$ for different $\Gamma$. 
Substituting $\sin\theta_{\Gamma}$ into Eqs.~(\ref{superfluid_density}) and (\ref{lambda}),  
we obtain~\cite{2017_Gurevich_Kubo}: 
\begin{eqnarray}
\frac{n_s(0, \Gamma, T)}{n_{s0}} 
= \frac{\lambda^{-2}(0, \Gamma, T)}{\lambda^{-2}_0} 
= \frac{2\Delta}{\pi}  {\rm Im} \,\psi \Biggl( \frac{1}{2} +\frac{\Gamma + i\Delta}{2\pi T}  \biggr) ,
\end{eqnarray}
Shown in Fig.~\ref{fig1} (c) are $n_s (0, \Gamma, T)$ and $\lambda^{-2}(0, \Gamma, T)$. 
The thermodynamic critical field $H_c$ is defined by $(\mu_0/2) H_c^2 = -\Omega (0, \Gamma, T)$, 
where the thermodynamic potential $\Omega$ is obtained by replacing $\omega_n$ in the BCS thermodynamic potential 
with $\omega_n + \Gamma$~\cite{2018_Herman}: 
\begin{eqnarray}
&& \hspace{-0.5cm} \Omega (0, \Gamma, T) = - 2\pi T N_0 \Delta \nonumber \\
&&\times  \sum_{\omega_n >0}  \biggl[ \frac{2(\omega_n + \Gamma)}{\Delta} (\cos \theta_{\Gamma} -1) + \sin\theta_{\Gamma} \biggr] . \label{Omega}
\end{eqnarray}
Shown in Fig.~\ref{fig1} (d) is $H_c$ as functions of $T$ for different $\Gamma$. 
As shown in Fig.~\ref{fig1} (b)-(d), 
$\Delta$, $n_s$, $\lambda^{-2}$, and $H_c$ are monotonically decreasing functions of $T$ and $\Gamma$.

\begin{figure}[tb]
   \begin{center}
   \includegraphics[height=0.47\linewidth]{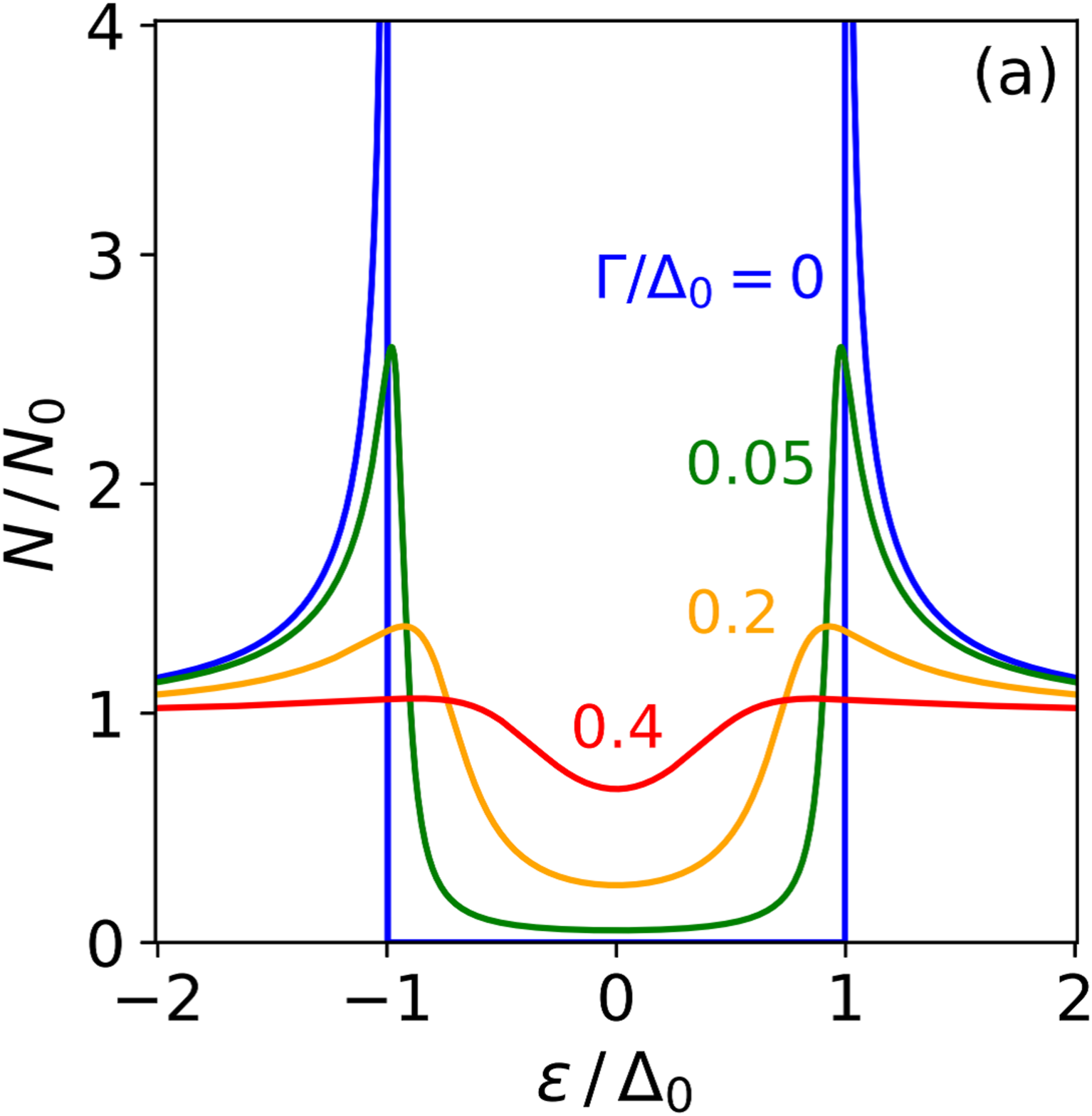}
   \includegraphics[height=0.47\linewidth]{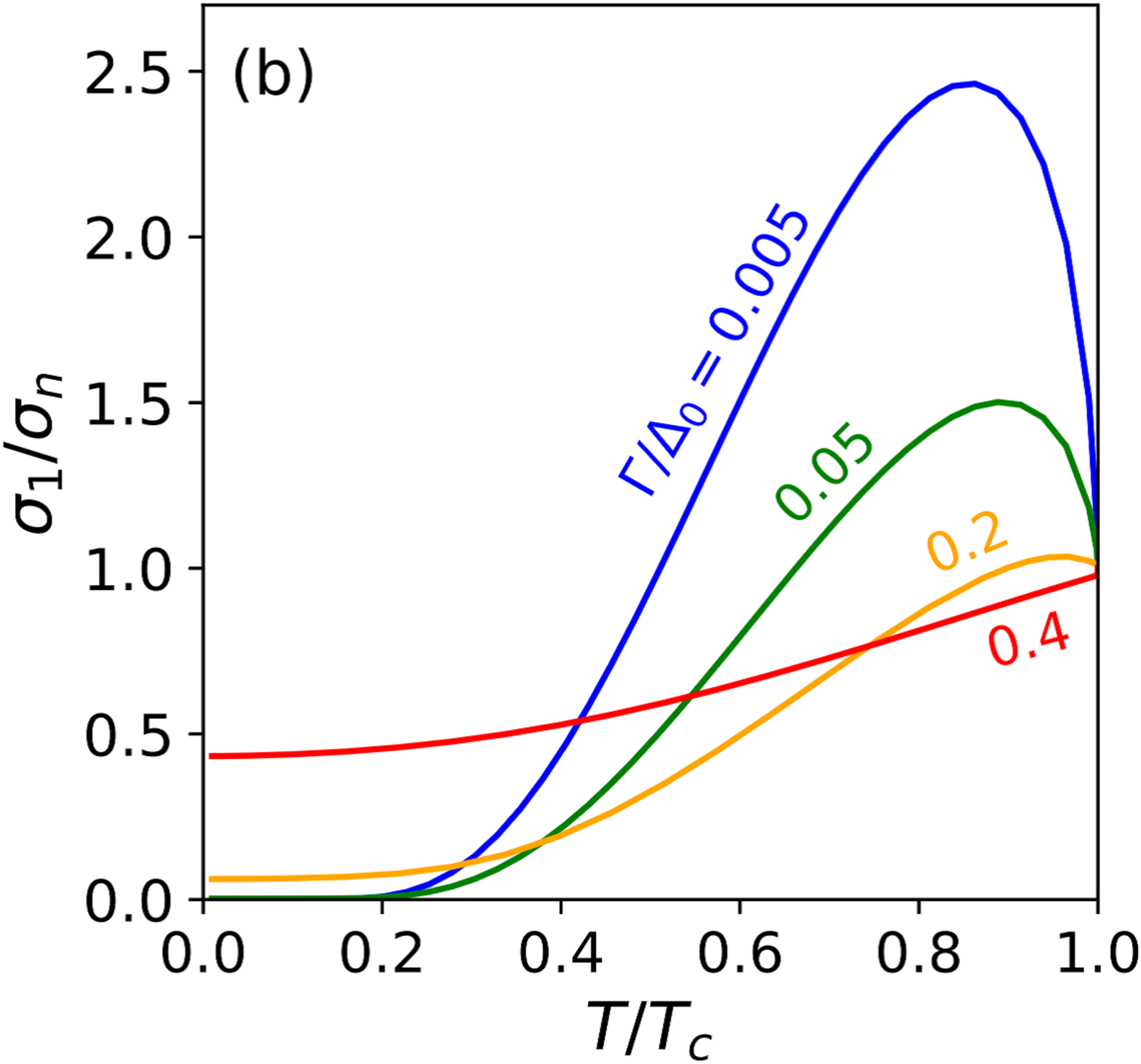}
   \includegraphics[width=0.49\linewidth]{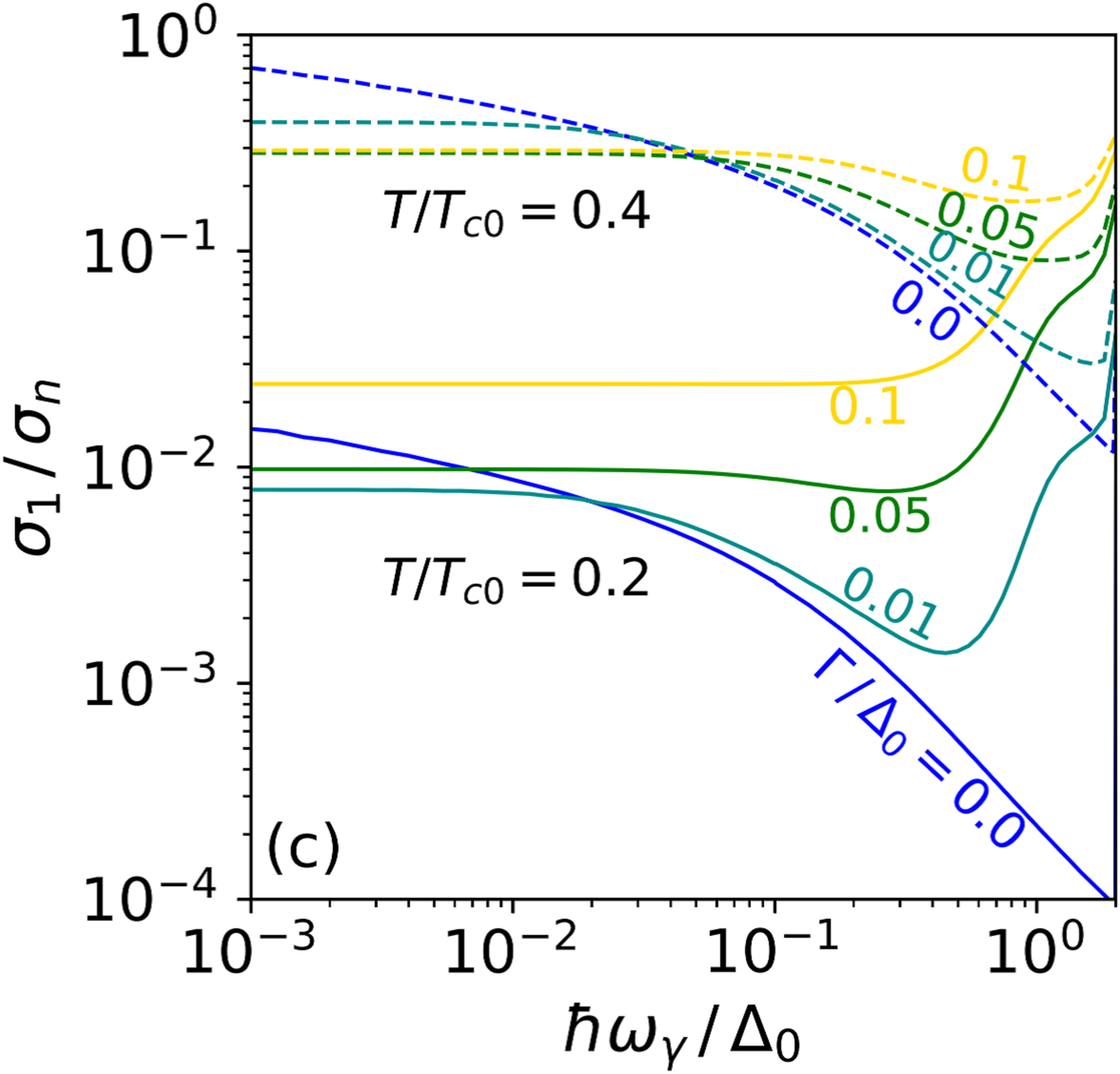}
   \includegraphics[width=0.49\linewidth]{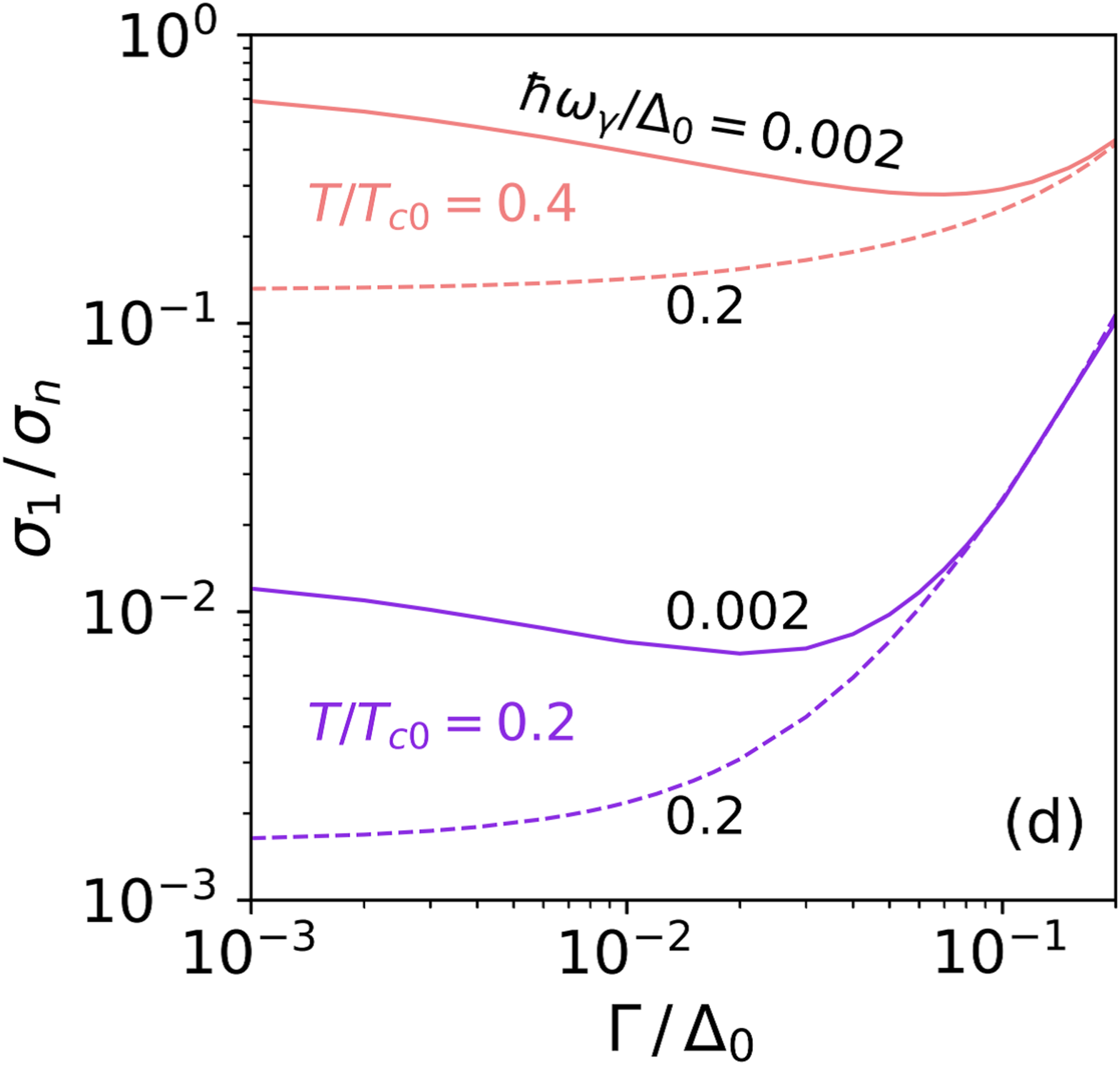}
   \end{center}\vspace{0 cm}
   \caption{
Quasiparticle DOS $N(\epsilon)$ and real part of complex conductivity $\sigma_1 (0, \Gamma, T, \omega_{\gamma})$  in the zero-current state. 
(a) $N(\epsilon)$ at $T/T_{c0}=0.1$. 
(b) $\sigma_1 (0, \Gamma, T, \omega_{\gamma})$ as functions of $T$, 
(c) $\omega_{\gamma}$, and
(d) $\Gamma$.   
   }\label{fig2}
\end{figure}

The retarded Green' functions are obtained by solving Eq.~(\ref{real_freq_Usadel}): 
$G^R = (\epsilon + i\Gamma)/\sqrt{(\epsilon + i\Gamma)^2 + \Delta^2}$ and $F^R = \Delta/\sqrt{(\epsilon + i\Gamma)^2 + \Delta^2}$. 
Then Eq.~(\ref{DOS}) reproduces the Dynes formula~\cite{2017_Gurevich_SUST, 2017_Gurevich_Kubo}
\begin{eqnarray}
\frac{N(\epsilon)}{N_0} = {\rm Re} G^R = {\rm Re} \frac{\epsilon + i\Gamma}{\sqrt{(\epsilon + i\Gamma)^2 -\Delta^2}} , 
\label{DynesDOS}
\end{eqnarray}
Shown in Fig.~\ref{fig2} (a) are the quasiparticle DOSs for different $\Gamma$.  
As $\Gamma$ increases, the DOS peaks are smeared out and the density of subgap states increases. 
At the Fermi level, the DOS is given by $N(0)/N_0 = \Gamma/\sqrt{\Gamma^2+\Delta^2}$ or $\simeq \Gamma/\Delta$ for $\Gamma \ll 1$.

Shown in Fig.~\ref{fig2} (b) are the $T$ dependences of $\sigma_1(0, \Gamma, T, \omega_{\gamma})$ for different $\Gamma$ calculated from Eq.~(\ref{sigma_1}). 
As $\Gamma$ increases, $\sigma_1$ decreases (increases) at higher (lower) $T$ regions and the coherence peak is suppressed. 
Shown in Fig.~\ref{fig2} (c) are the $\omega_{\gamma}$ dependences of $\sigma_1$ for different $\Gamma$ at $T/T_{c0}=0.2$ (solid curves) and $T/T_{c0}=0.4$ (dashed curves). 
It is clearly shown that $\sigma_1$ is determined by $\Gamma$ rather than $\omega_{\gamma}$ for $\omega_{\gamma} < \Gamma$. 
As a result, the divergence at $\omega_{\gamma}\to 0$ disappears. 
The rapid increase of $\sigma_1$ at $\omega_{\gamma} \simeq 2\Delta$ is the photon absorption edge. 
For a finite $\Gamma$, the second edge appears at $\omega_{\gamma} \simeq \Delta$ due to the finite density of subgap states. 
Shown in Fig.~\ref{fig2} (d) are the $\Gamma$ dependences of $\sigma_1$ at $\omega_{\gamma}=0.002$ (solid curves) and $\omega_{\gamma}=0.2$ (dashed curves). 
A finite $\Gamma$ can reduce $\sigma_1$ for $\omega_{\gamma} \ll T$ (solid curves), 
while increases $\sigma_1$ for $\omega_{\gamma} \gtrsim T$ (dashed curves). 
These results can be summarized as follows.  
At low temperatures $T < (\omega_{\gamma}, \Gamma)$ for which $\sigma_1$ is dominated by quasiparticles with $\epsilon \ll \Delta$, 
a finite DOS at the Fermi level increases $\sigma_1$, 
giving rise to a residual conductivity $\sigma_1/\sigma_n \to \Gamma^2/(\Gamma^2 + \Delta^2)$~\cite{2017_Gurevich_SUST, 2017_Gurevich_Kubo, 2017_Herman}. 
At $T \gg (\omega_{\gamma}, \Gamma)$, 
where $\sigma_1$ is mostly determined by thermally activated quasiparticles, 
the broadening of the DOS due to a finite $\Gamma$ reduces $\sigma_1$~\cite{2017_Gurevich_SUST, 2014_Gurevich, 2017_Gurevich_Kubo}. 
The reduction of $\sigma_1$ can be qualitatively understood from the similar discussion as in Section I. 
The convolution of the BCS DOS $N(\epsilon)$ and $N(\epsilon+\omega_{\gamma})$ yields the logarithmic factor which diverges at $\omega_{\gamma}\to 0$ in the MB formula. 
When $\Gamma > \omega_{\gamma}$, 
the denominator in the logarithmic factor is replaced with $\Gamma$ 
and the divergence at $\omega_{\gamma}\to 0$ disappears. 
As $\Gamma$ increases, $\sigma_1$ logarithmically decreases~\cite{2017_Gurevich_SUST, 2014_Gurevich, 2017_Gurevich_Kubo}.

\section{Current-carrying state} \label{current_carrying_state}


\subsection{Pair potential, superfluid density, penetration depth, and supercurrent density} 

\begin{figure}[tb]
   \begin{center}
   \includegraphics[width=0.48\linewidth]{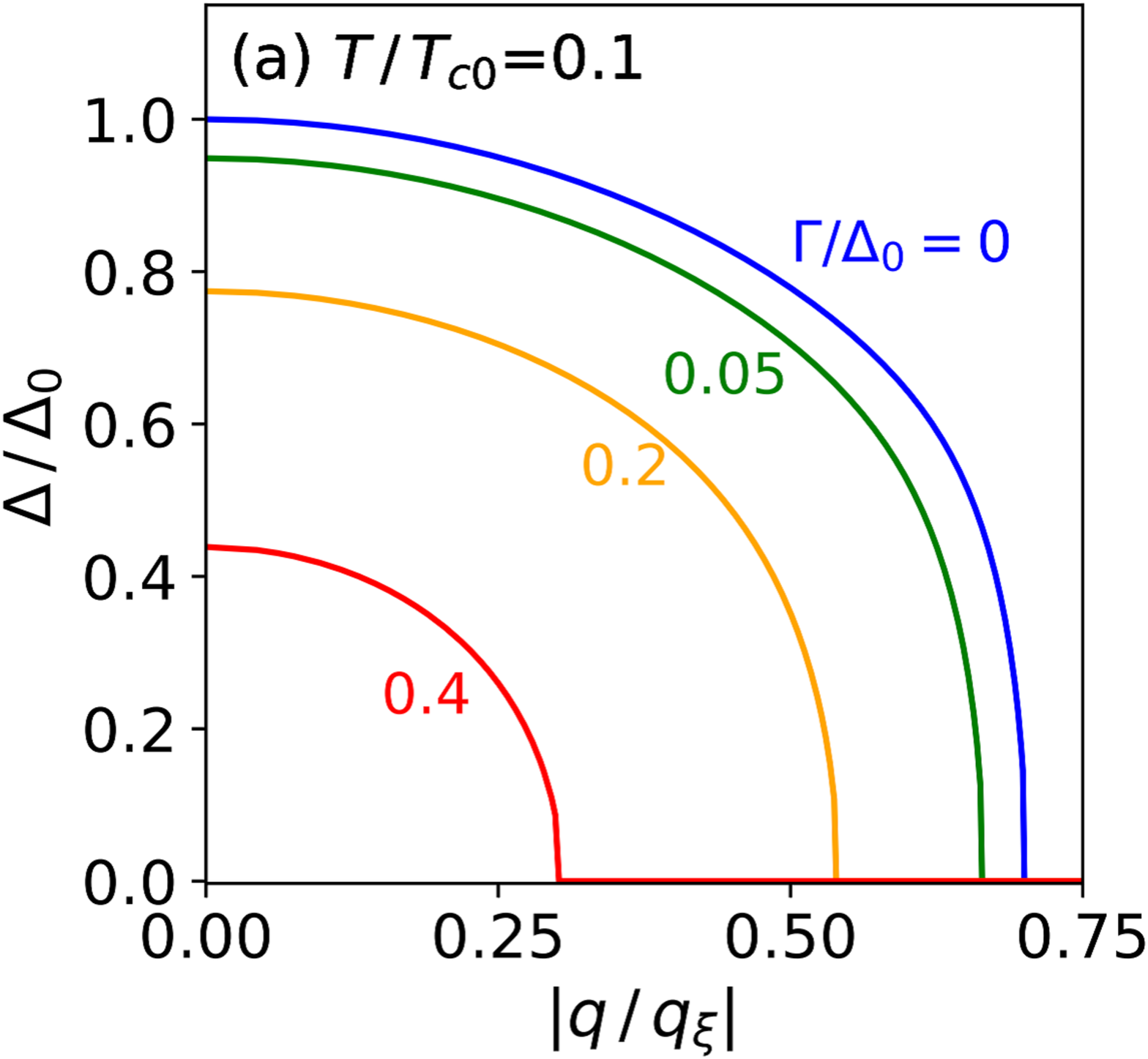}
   \includegraphics[width=0.48\linewidth]{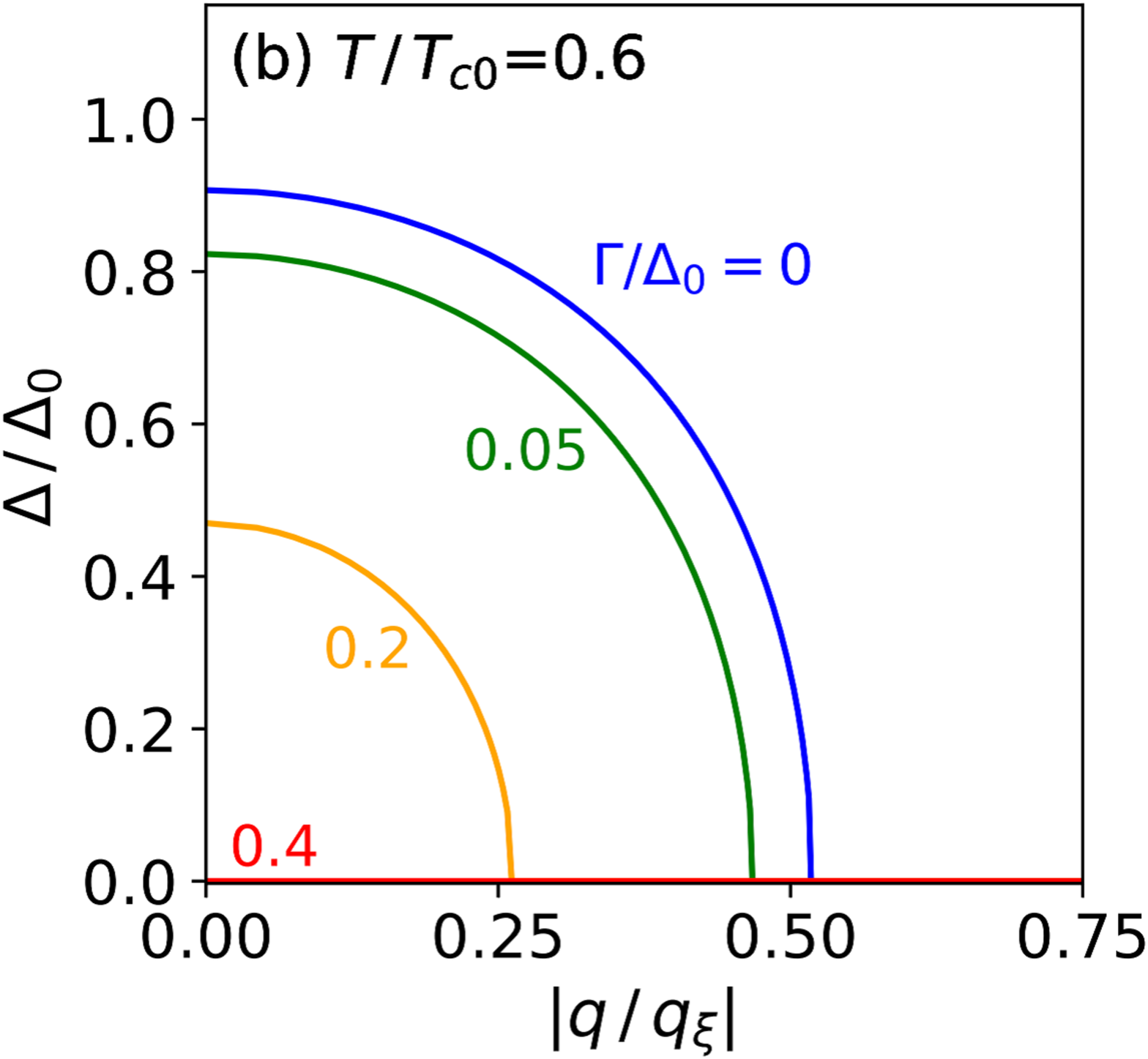}
   \includegraphics[width=0.48\linewidth]{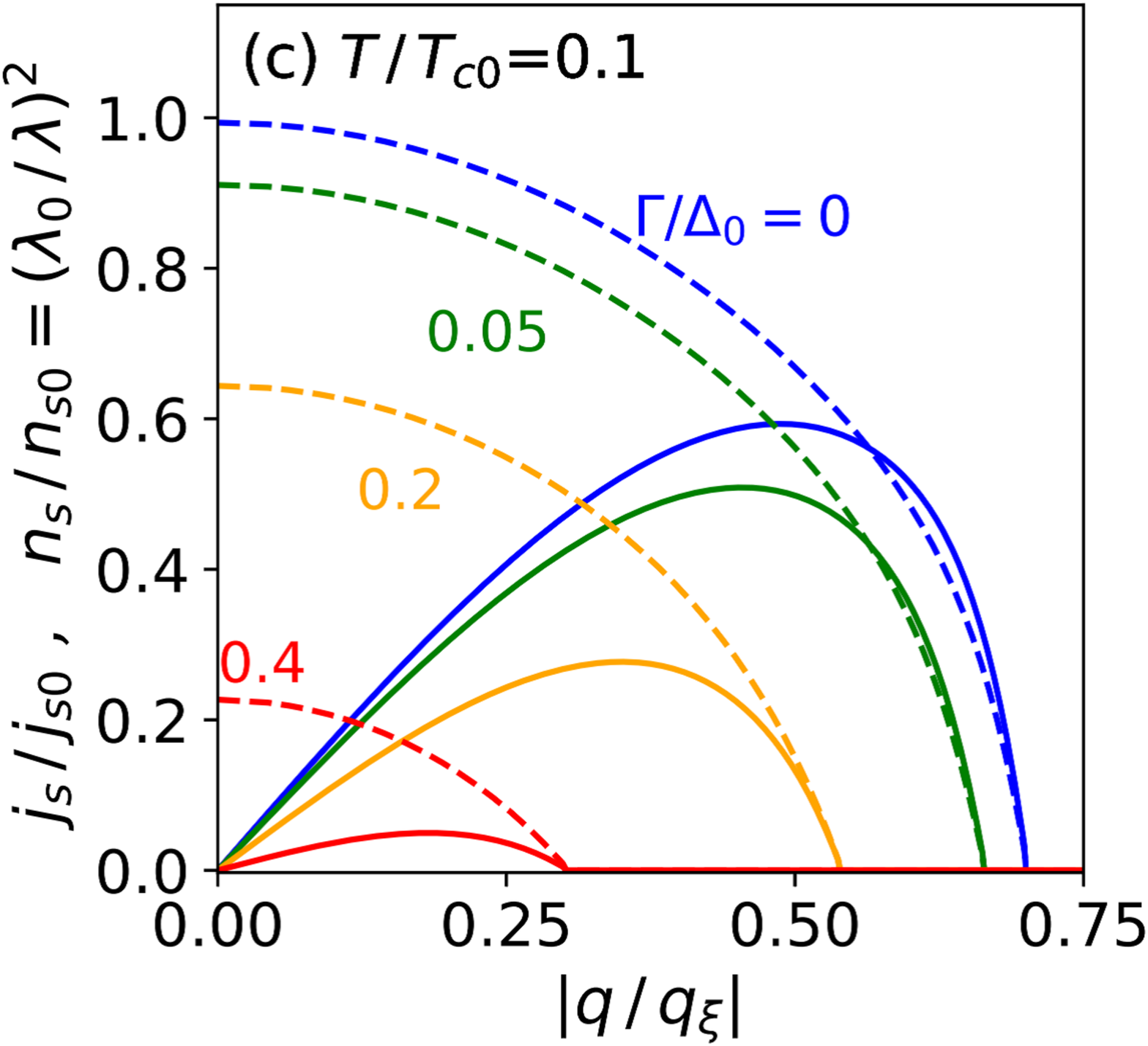}
   \includegraphics[width=0.48\linewidth]{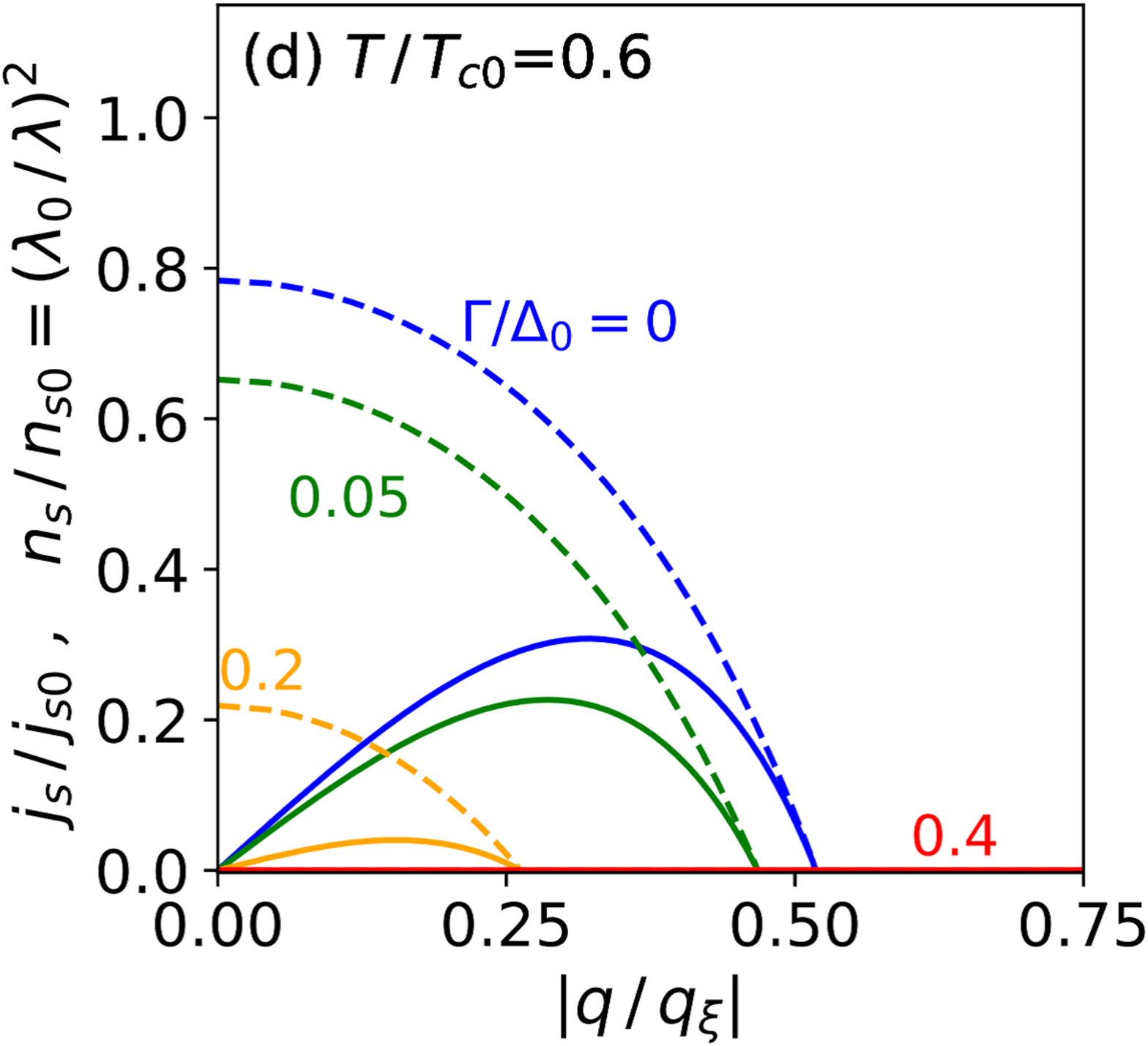}
   \end{center}\vspace{0 cm}
   \caption{
(a, b) Pair potential $\Delta$ as functions of superfluid momentum $|q/q_{\xi}|=\sqrt{s/\Delta_0}$ for different $\Gamma$ and $T$. 
(c, d) Superfluid density $n_s$ (dashed curves) and supercurrent density $j_s$ (solid curves) as functions of  $|q|$ for different $\Gamma$ and $T$. 
The peak value of $j_s$ is the depairing current density $j_d(\Gamma, T)$.  
   }\label{fig3}
\end{figure}

Now consider current-carrying states ($s \propto q^2 \ne 0$). 
The pair potential $\Delta = \Delta (s, \Gamma, T)$ is obtained by solving Eqs.~(\ref{thermodynamic_Usadel}) and (\ref{self-consistency}). 
For a special case $(s, \Gamma, T/T_c) \ll 1$, 
by setting $\theta = \theta_{\Gamma}+\delta \theta$ and $\Delta = \Delta_{\Gamma} + \delta\Delta$ and linearizing Eqs.~(\ref{thermodynamic_Usadel}) and (\ref{self-consistency}), 
we obtain a convenient formula $\Delta (s, \Gamma, 0) = 1 - \Gamma - (\pi/4) s$. 
For a general set of $s$, $\Gamma$, and $T$, 
we need to numerically solve Eqs.~(\ref{thermodynamic_Usadel}) and (\ref{self-consistency}) or, in the more convenient forms, 
$(\Delta - s/\sqrt{1+ z^2} ) z = \omega_n + \Gamma $ and
$\Delta \ln (T_{c0}/T) = 2\pi T \sum_{n} ( \Delta/\omega_n - 1/\sqrt{1+z^2} )$. 
Here $z = \cot\theta$. 
Shown in Fig.~\ref{fig3} (a) and (b) are the pair potential $\Delta$ as functions of the superfluid momentum $|q|$ for different $\Gamma$ and $T$.  
The blue curves ($\Gamma=0$) represents $\Delta$ for the ideal dirty BCS superconductors~\cite{1963_Maki_I, 1963_Maki_II}. 
The other curves exhibit smaller $\Delta$ due to the pair-breaking effect of $\Gamma>0$. 
As $s\, (\propto q^2)$, $\Gamma$, or $T$ increase, $\Delta$ monotonically decreases.

Shown in Fig.~\ref{fig3} (c) and (d) are the superfluid density, penetration depth, and supercurrent density as functions of $|q|$ for different $\Gamma$ and $T$ calculated from Eqs.~(\ref{superfluid_density})-(\ref{supercurrent}). 
The superfluid density $n_s$ and penetration depth $\lambda^{-2}$ (dashed curves) are monotonically decreasing functions of $\Gamma$, $|q|$, and $T$, 
but the supercurrent density $j_s$ (solid curves) exhibits non-monotonic behaviors. 
At smaller $|q|$ regions, $j_s$ increases with $|q|$. 
However, when $|q|$ reaches a critical value $q_d(\Gamma, T)$, 
$j_s$ ceases to increase because of a rapid reduction of superfluid density $n_s$ at higher $|q|$ regions.  
The maximum value of $j_s$ is the so-called depairing current density $j_d$. 
The solid blue curves ($\Gamma =0$) reproduce the well-known results for the ideal dirty BCS superconductors~\cite{1963_Maki_I, 1963_Maki_II, 1980_Kupriyanov, 2012_Clem_Kogan}. 
The other solid curves ($\Gamma>0$) show that both $q_d$ and $j_d$ decrease as $\Gamma$ increases.

\subsection{Depairing current density} 

\begin{figure}[tb]
   \begin{center}
   \includegraphics[width=0.48\linewidth]{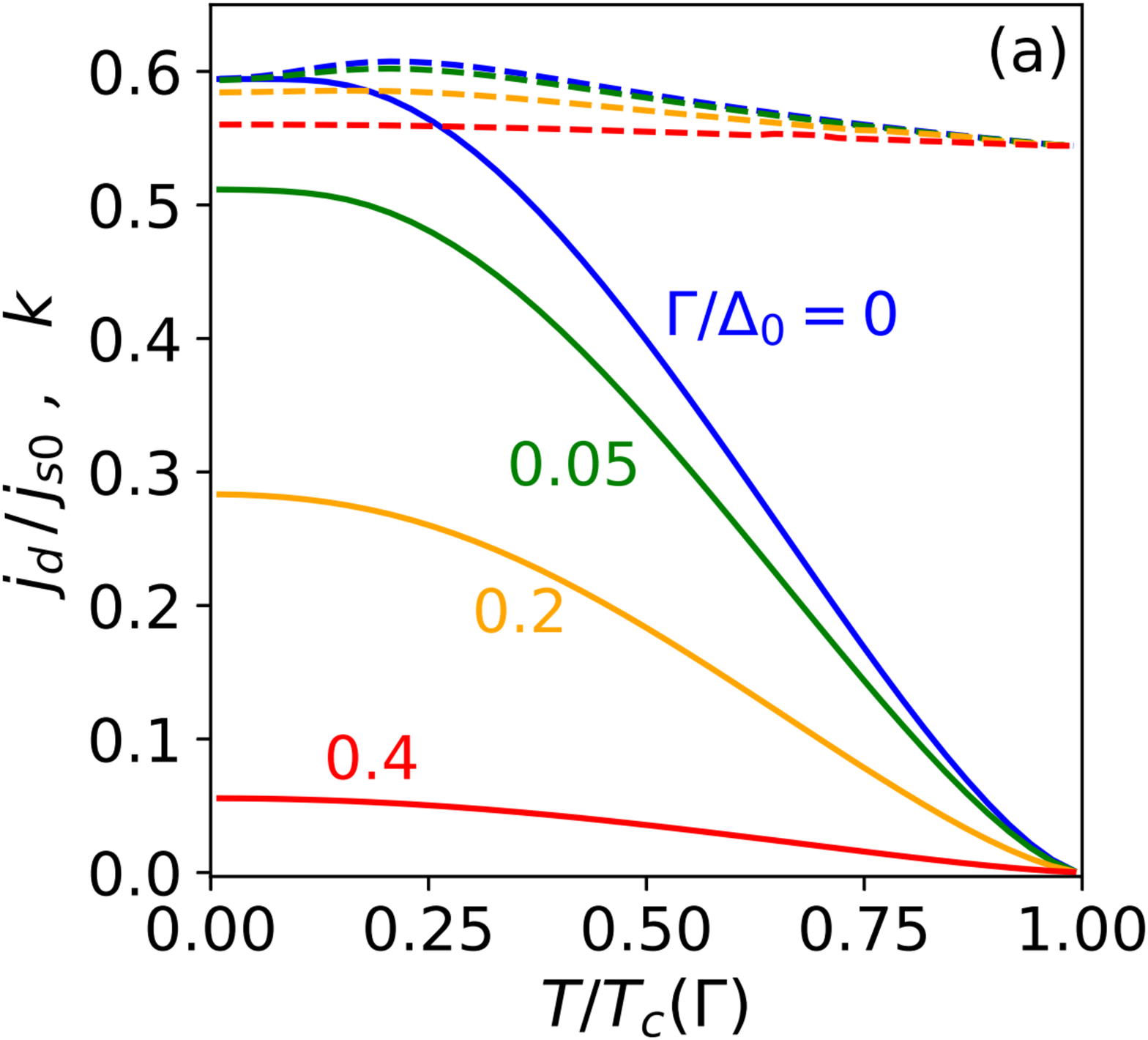}
   \includegraphics[width=0.48\linewidth]{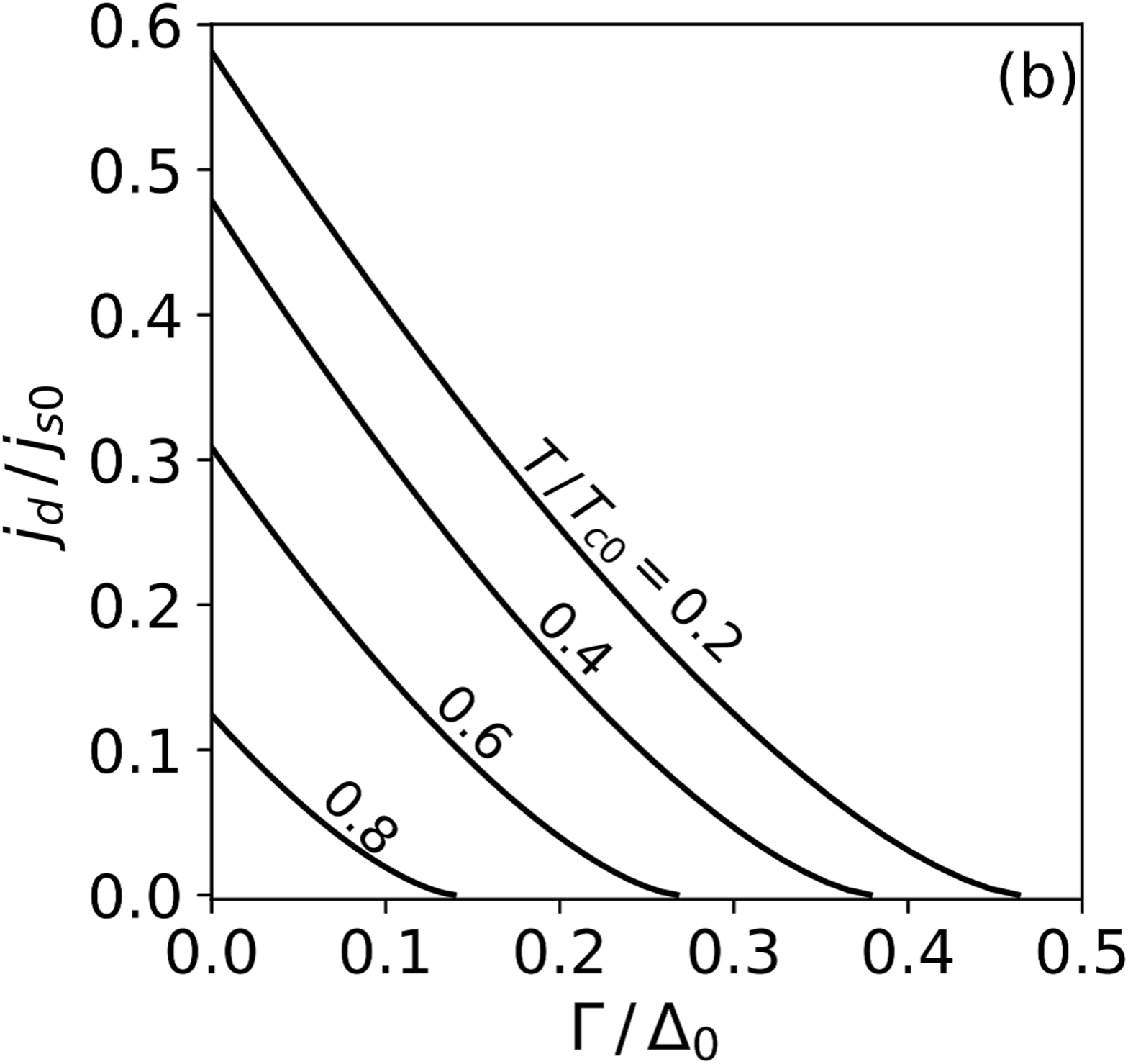}
   \includegraphics[width=0.48\linewidth]{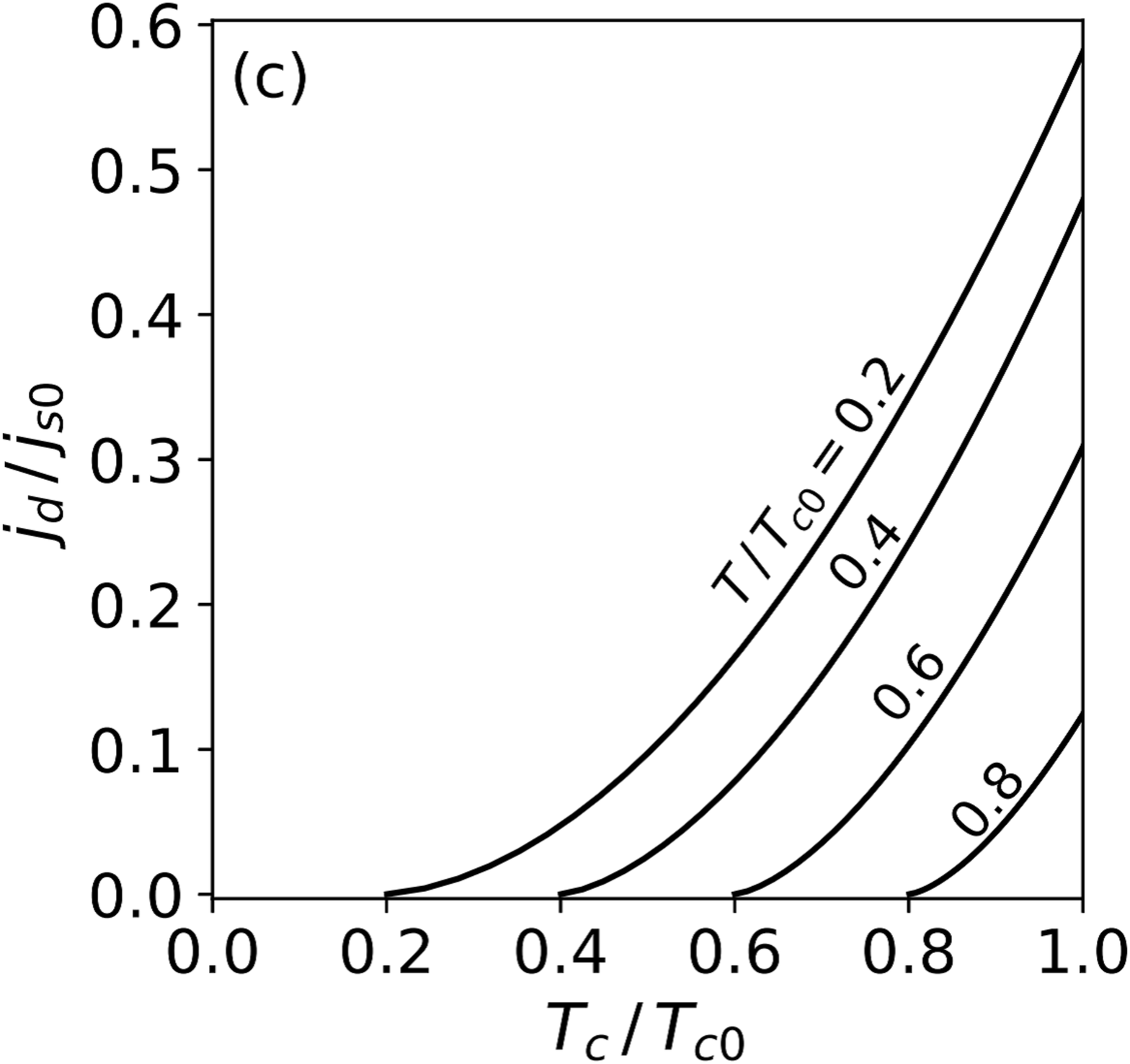}
   \includegraphics[width=0.48\linewidth]{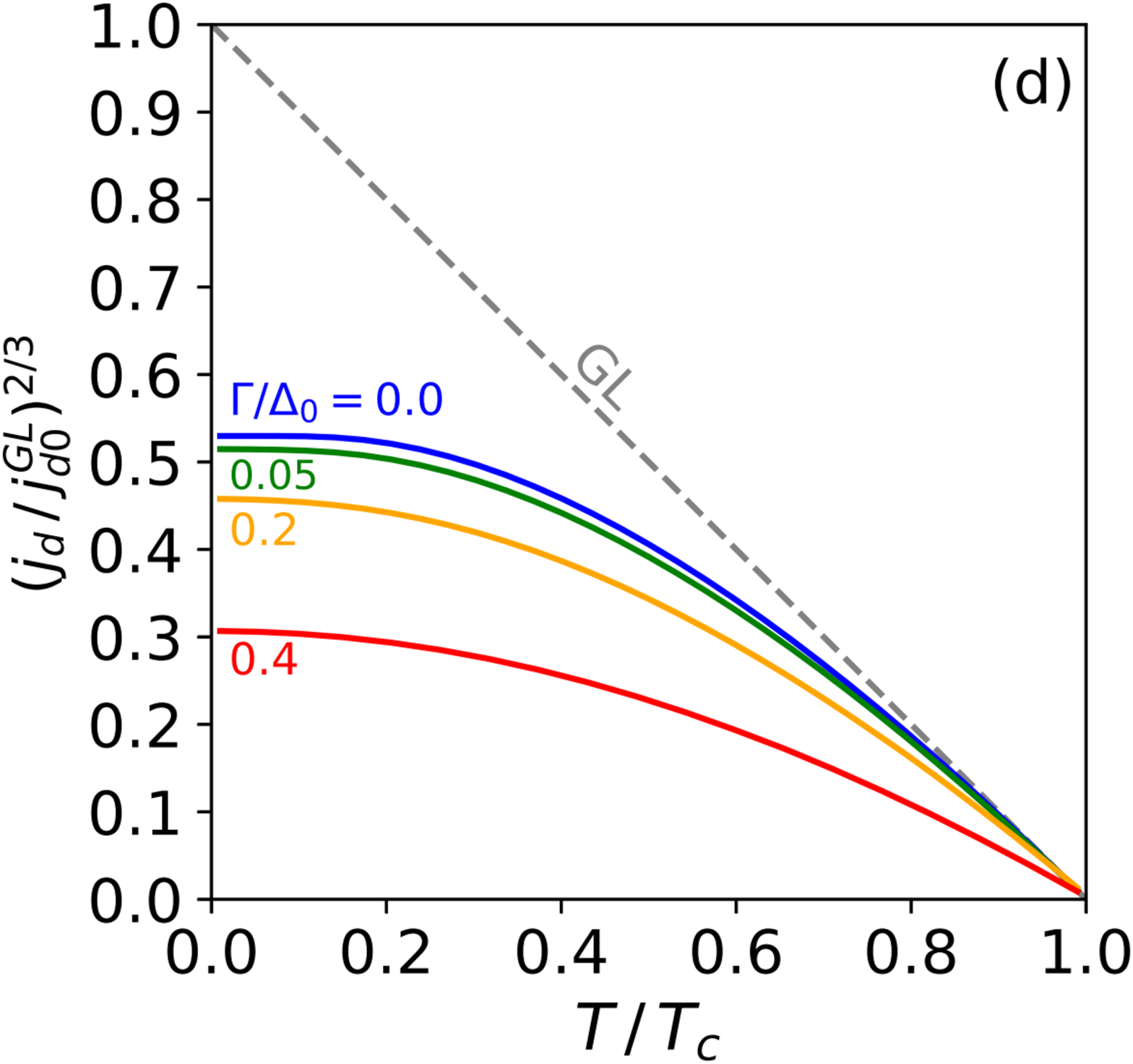}
   \end{center}\vspace{0 cm}
   \caption{
(a) Depairing current density $j_d$ (solid curves) and the $k$ parameter (dashed curves) as functions of $T$ for different $\Gamma$. 
The definition of $k$ is given by Eq.~(\ref{k}). 
(b) $j_d$ as functions of $\Gamma$ and (c) functions of $T_c$. 
(d) $(j_d/j_{d0}^{\rm GL})^{2/3}$ as functions of $T/T_c$. 
Here the normalization factor $j_{d0}^{\rm GL}=j_{d}^{\rm GL}(\Gamma,0)$ is given by Eq.~(\ref{jd0GL}). 
The GL result extrapolated to $T\ll T_c$ is also shown for comparison (dashed gray line). 
   }\label{fig4}
\end{figure}

Here we discuss the depairing current density $j_d (\Gamma, T)$ more in details.  
The solid curves in Fig.~\ref{fig4} (a) are $j_d$ as functions of $T$ for different $\Gamma$. 
The solid blue curve ($\Gamma=0$) corresponds to $j_d$ for the ideal dirty BCS superconductors, 
which takes the maximum value $j_d (0,0) = 0.595 H_{c0}/\lambda_0$ consistent with the previous study by Kupriyanov and Lukichev~\cite{1980_Kupriyanov, 2012_Clem_Kogan}. 
The other solid curves ($\Gamma >0$) yield smaller $j_d$ than the ideal case due to the $\Gamma$-induced degradation of $n_s$.  
Shown in Fig.~\ref{fig4} (b) and Fig.~\ref{fig4} (c) are $j_d$ as functions of $\Gamma$ and $T_c(\Gamma)$, respectively, for various temperatures.  
As $\Gamma$ increases (as $T_c$ decreases), $j_d$ monotonically decreases.

It is sometimes convenient to express $j_d$ as
\begin{eqnarray}
j_d (\Gamma, T) = k \frac{H_c(\Gamma, T)}{\lambda(0, \Gamma, T)} , \label{k}
\end{eqnarray}
Here $k$ is a coefficient. 
Since $\lambda (0, \Gamma, T)$, $H_c(\Gamma, T) $, and $j_d(\Gamma,T)$ are already calculated in Figs.~\ref{fig1} (c), (d) and Fig.~\ref{fig4} (a), respectively, 
it is straightforward to calculate the coefficient $k$.  
The dashed curves in Fig.~\ref{fig4} (a) are $k$ as functions of $T$ for different $\Gamma$. 
The dashed blue curve ($\Gamma=0$) corresponds to $k$ for the ideal dirty BCS superconductors, 
consistent with the previous studies (see e. g., Ref.~\cite{2012_Clem_Kogan}). 
The other dashed curves ($\Gamma>0$) exhibit different $T$ dependences from the ideal dirty BCS superconductors with $\Gamma=0$, 
but all the curves merge to the well-known Ginzburg-Landau (GL) result 
$k = 2\sqrt{2}/3\sqrt{3}=0.544$ at $T \simeq T_c$ independent of $\Gamma$.

To understand the behavior at $T\simeq T_c$, 
we derive the GL equation for superconductors with a finite $\Gamma$. 
For $T$ close to $T_c$, the pair potential $\Delta$ becomes small, and we can expand the thermodynamic Green's functions in powers of $\delta = \Delta/2\pi T \ll 1$. 
Substituting $F=\sin\theta =  \sum_m F_m \delta^m$ 
and $G= \sqrt{1-F(\delta)^2} = \sum_m (1/m!)(d^m G/d\delta^m) \delta^m$ into Eq.~(\ref{thermodynamic_Usadel}), we identify $F_m$: 
\begin{eqnarray}
&&\sin\theta = F_1 \delta - \frac{\delta^3}{2} (F_1^3 - \bar{s} F_1^4) , \\
&&\cos\theta = 1 -\frac{\delta^2}{2} F_1^2 - \frac{\delta^4}{8} (4\bar{s} F_1^5 - 3 F_1^4)
\end{eqnarray}
Here $F_1=\delta/(n+1/2+ s/2\pi T +\Gamma/2\pi T)$. 
Then Eq.~(\ref{self-consistency}) yields 
$\ln (T_{c0}/T)=(\pi/4T)(s+\Gamma) + (7\zeta(3)/8\pi^2 T^2) \Delta^2$. 
Subtracting the equation for $T_c$, $\ln (T_{c0}/T_{c}) = \pi \Gamma/4T_{c}$, 
we obtain the GL equation for the Dynes model
\begin{eqnarray}
1 - \frac{T}{T_{c}} = \frac{\pi s}{4 T_{c}} + \frac{7\zeta(3)}{8\pi^2 T_{c}^2} \Delta^2 , \label{GL}
\end{eqnarray}
for $\Delta, \, s, \, \Gamma \ll 2\pi T$ and $T \simeq T_{c}(\Gamma)$. 
This has the similar form as the well-known GL equation. 
The only difference is that $T_{c0}$ is replaced with $T_c(\Gamma)$. 
So, obviously, Eq.~(\ref{GL}) should yield the well-known GL depairing current density independent of $\Gamma$. 
The solution of Eq.~(\ref{GL}) is
\begin{eqnarray}
\Delta (s, \Gamma, T) 
= \sqrt{ \frac{8\pi^2 T_{c}(\Gamma)^2}{7\zeta(3)} 
\biggl(1- \frac{T}{T_{c}(\Gamma)} \biggr)  \biggl( 1-\frac{s}{s_m(\Gamma, T)} \biggr) }, 
\nonumber \\
\end{eqnarray}
where $s_m(\Gamma,T) = (4T_{c}/\pi) (1-T/T_{c})$. 
Then Eqs.~(\ref{superfluid_density}), (\ref{lambda}) and Eq.~(\ref{Omega}) yield 
\begin{eqnarray}
&&\frac{n_s (s, \Gamma, T)}{n_{s0}} 
= \frac{\lambda^{-2} (s, \Gamma, T)}{\lambda_0^{-2}} 
=\frac{\Delta^2 (s, \Gamma, T)}{2T_{c}} , \\
&&H_c (\Gamma, T) 
= \sqrt{ \frac{8\pi^2 T_{c}^2 N_0}{7\zeta(3) \mu_0}} \biggl( 1- \frac{T}{T_{c}} \biggr)
\end{eqnarray}
at $T\simeq T_{c}(\Gamma)$. 
Then Eq.~(\ref{supercurrent}) yields
\begin{eqnarray}
j_s (s, \Gamma, T) 
= \sqrt{ \frac{\pi}{2(T_{c}-T)} } \sqrt{s} \biggl( 1- \frac{s}{s_m} \biggr) \frac{H_c(\Gamma, T)}{\lambda(0, \Gamma, T)} . 
\end{eqnarray}
This takes the maximum when $s = s_m/3$. 
Thus, the depairing current density at $T \simeq T_{c}(\Gamma)$ is given by
\begin{eqnarray}
j_d^{\rm GL} (\Gamma, T) = j_s (s_m/3, \Gamma, T) 
= \frac{2\sqrt{2}}{3\sqrt{3}} \frac{H_c(\Gamma, T)}{\lambda(0, \Gamma, T)} . \label{jd_GL_1}
\end{eqnarray}
As expected, the coefficient $k$ corresponds with the well-known GL result independent of $\Gamma$ at $T \simeq T_c$. 
Eq.~(\ref{jd_GL_1}) can be rewritten as
\begin{eqnarray}
j_d^{\rm GL} (\Gamma, T) =\frac{16 j_{s0}}{21 \zeta(3)} \sqrt{\frac{\pi}{3}} 
\biggl( \frac{e^{\gamma_E}T_c}{T_{c0}} \biggr)^{\frac{3}{2}} \biggl( 1 - \frac{T}{T_c} \biggr)^{\frac{3}{2}} \label{jd_GL_2} , 
\end{eqnarray}
yielding the well-known $T$ dependence in the GL regime.

Measurements of $j_d$ are often summarized by plotting $(j_d/j_{d0}^{\rm GL})^{2/3}$ as functions of $T/T_c$ (see e.g. Refs.~\cite{1982_Romijn, 2004_Rusanov}). 
Here the normalization constant is given by 
\begin{eqnarray}
j_{d0}^{\rm GL} 
&=& j_d^{\rm GL}(\Gamma , 0)= 1.54 j_{s0} \biggl( \frac{T_c}{T_{c0}} \biggr)^{\frac{3}{2}} \nonumber \\
&=& \frac{8\pi^2\sqrt{2\pi}}{21\zeta(3) e} \sqrt{\frac{(k_B T_c)^3}{\hbar v_F \rho (\rho \ell)}} .
\label{jd0GL}
\end{eqnarray}
The solid curves in Fig.~\ref{fig4} (d) are our theoretical results valid at an arbitrary temperature $0\le T \le T_c$. 
The solid blue curve ($\Gamma=0$) is coincident with the well-known Kupriyanov-Lukichev curve~\cite{1980_Kupriyanov}, 
which reaches $(j_d/j_{d0}^{\rm GL})^{2/3}=0.53$ at $T\to 0$. 
The other solid curves represent $j_d$ for $\Gamma>0$, 
in which the deviations from the Kupriyanov-Lukichev curve increases with $\Gamma$. 
The dashed gray line represent the GL result, which is valid at $T\simeq T_c$.

\subsection{Density of states} 

\begin{figure}[tb]
   \begin{center}
   \includegraphics[width=0.48\linewidth]{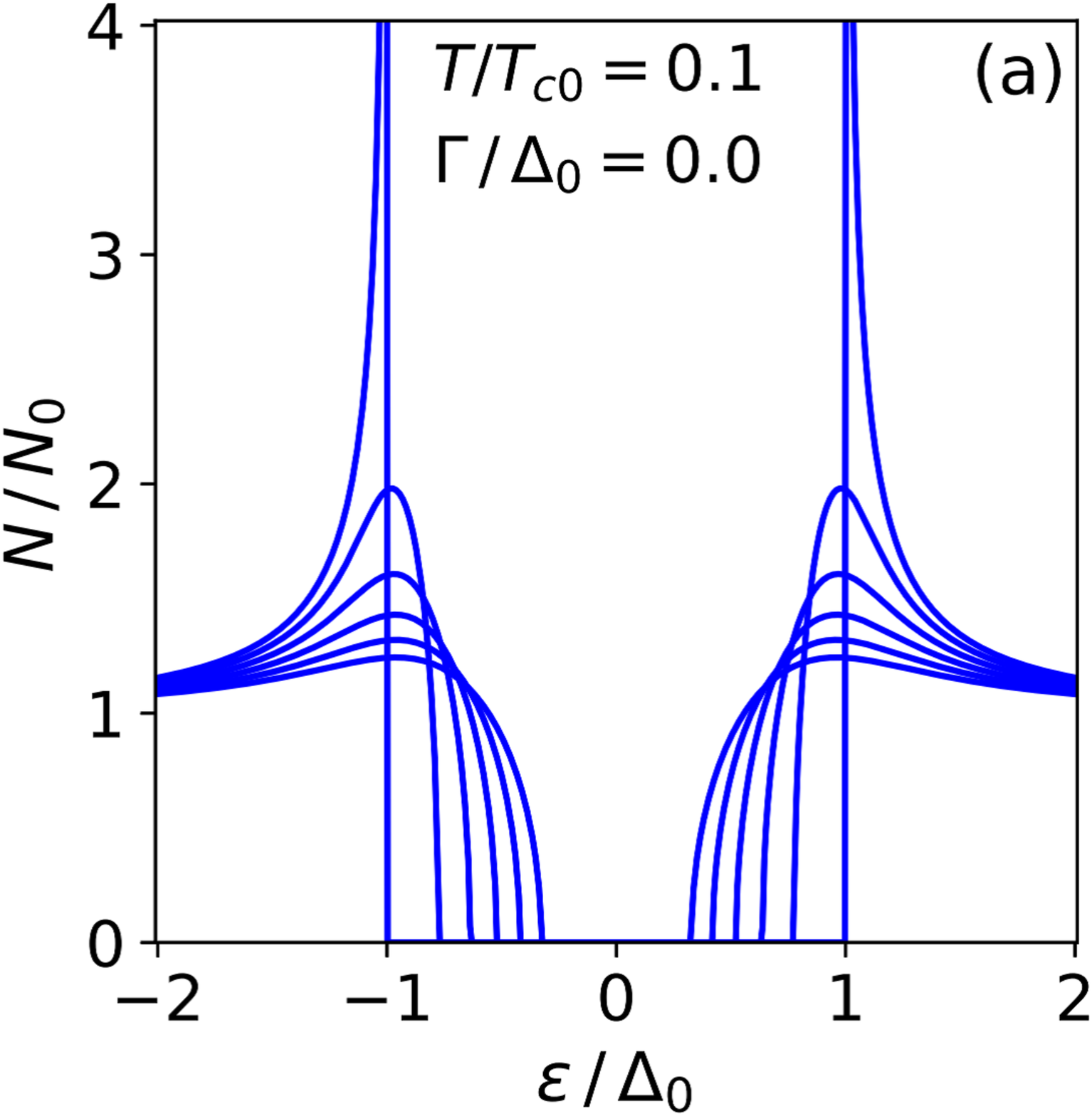}
   \includegraphics[width=0.48\linewidth]{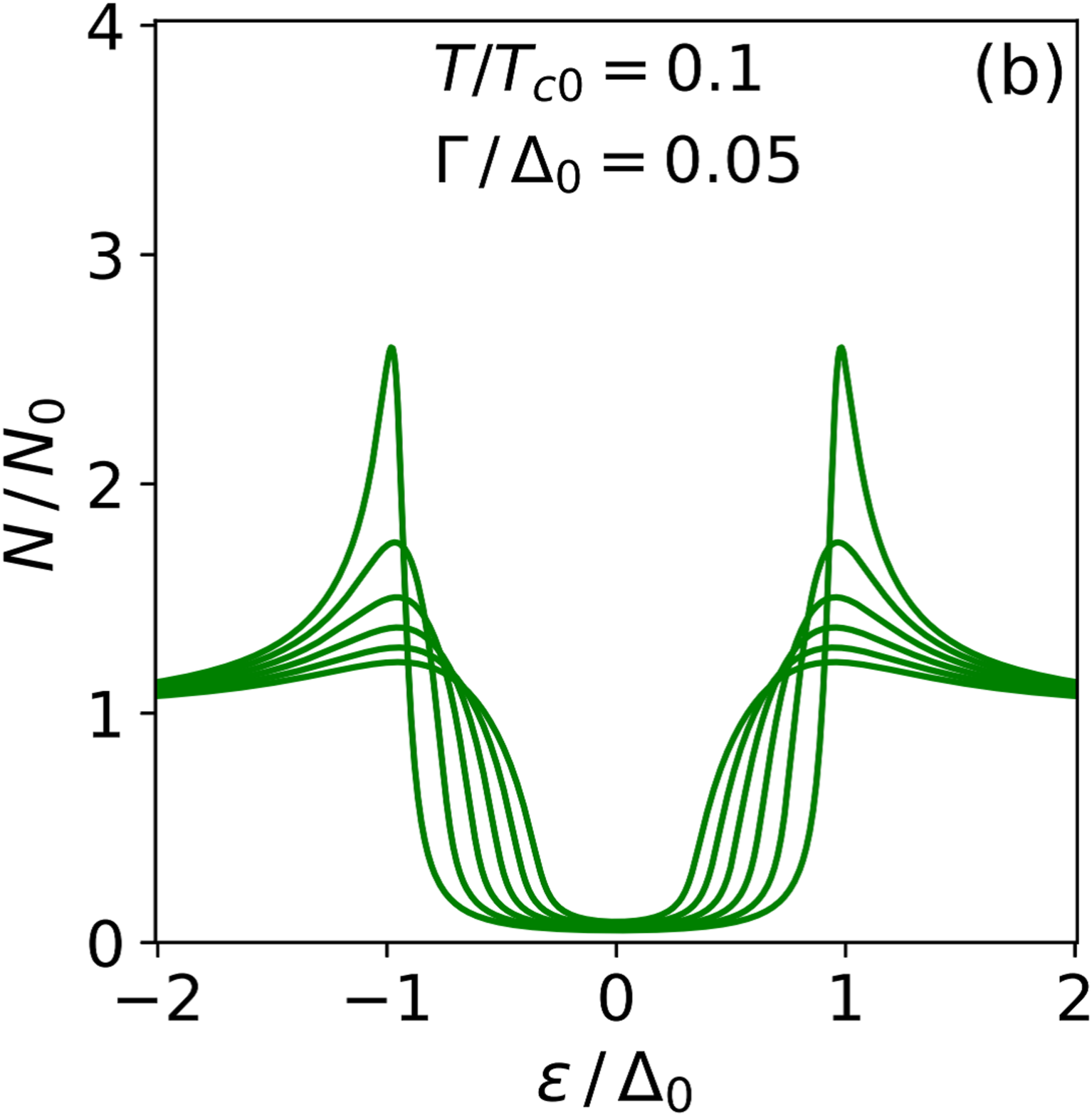}
   \includegraphics[width=0.48\linewidth]{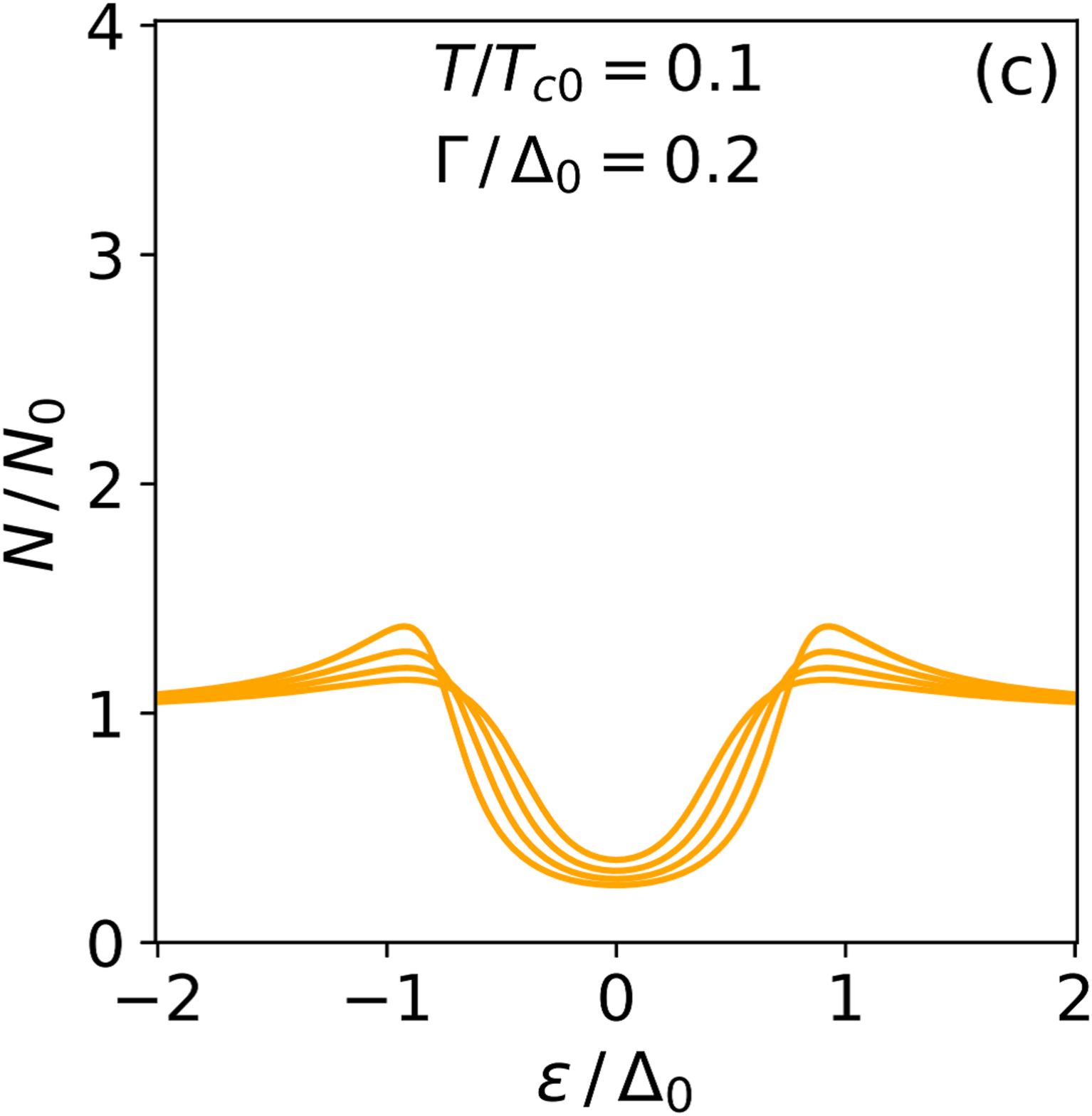}
   \includegraphics[width=0.48\linewidth]{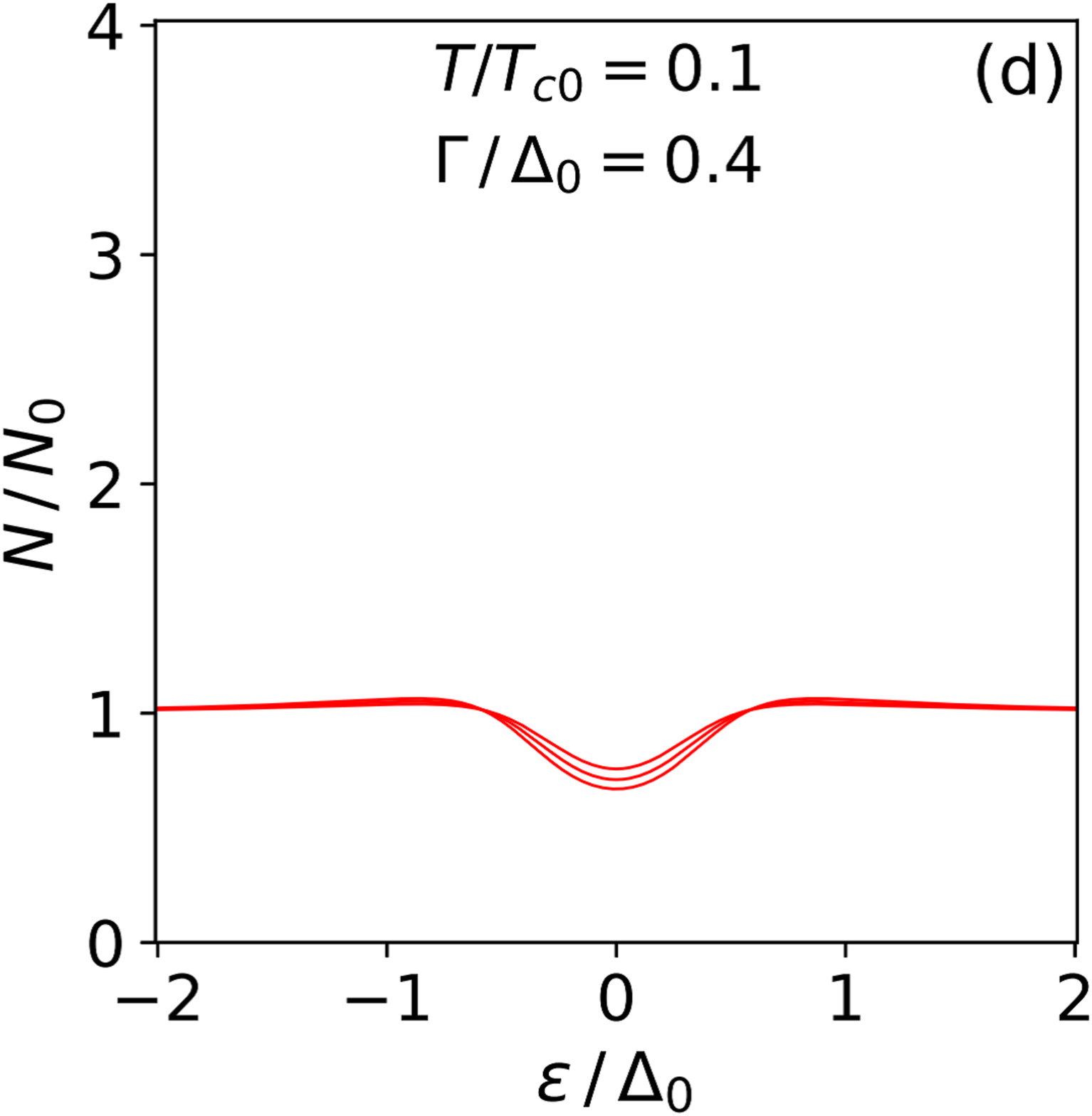}
   \end{center}\vspace{0 cm}
   \caption{
Quasiparticle DOS calculated from Eq.~(\ref{DOS}) for (a) $\Gamma=0$, (b) $0.05$, (c) $0.2$, and (d) $0.4$. 
Curves with the highest (lowest) peaks correspond to $j_s=0$ ($j_s= j_d$).  
   }\label{fig5}
\end{figure}

Now we solve the real-frequency Usadel equation. 
Substituting $\Delta$ obtained in the above (see Fig.~\ref{fig3}) into Eq.~(\ref{real_freq_Usadel}), 
we can calculate the retarded Green's functions $G^R$ and $F^R$. 
Then the quasiparticle DOS is given by Eq.~(\ref{DOS}). 
Shown in Fig.~\ref{fig5} (a) are the effects of pair-breaking currents on the quasiparticle DOS for the ideal dirty BCS superconductors with $\Gamma=0$~\cite{1964_Maki_current}. 
The curve with the highest peak represents the zero-current state ($j_s=0$). 
As $j_s$ increases, the singularity in the ideal BCS DOS disappears and the DOS peaks are broadened. 
The curve with the lowest peak represents the DOS under the depairing current ($j_s=j_d$). 
Note here we have the gapped spectrum even at $j_s=j_d$,
which is the characteristic of dirty or moderately dirty superconductors.  
In clean superconductors, the spectrum gap disappears before reaching $j_d$~\cite{2012_Lin_Gurevich}. 
Shown in Fig.~\ref{fig5} (b)-(d) are the effects of the current on DOS for $\Gamma > 0$. 
Even for the zero-current states (the curves with the highest peaks), 
the DOS peaks are smeared out by the pair-breaking $\Gamma$ as seen in Fig.~\ref{fig1} (e). 
As the current increases, 
the DOS peaks are even more broadened and the density of subgap states increases. 
For instance, the DOS at $\epsilon=0$ is given by
$N(0)/N_0=(\Gamma/\Delta)[1+(s/\Delta)(1+\pi/4)]$ for $(s, \Gamma) \ll 1$~\cite{2019_Kubo_Gurevich}.

\subsection{Dissipative conductivity $\sigma_1$ under a dc bias} 

\begin{figure}[tb]
   \begin{center}
   \includegraphics[width=0.494\linewidth]{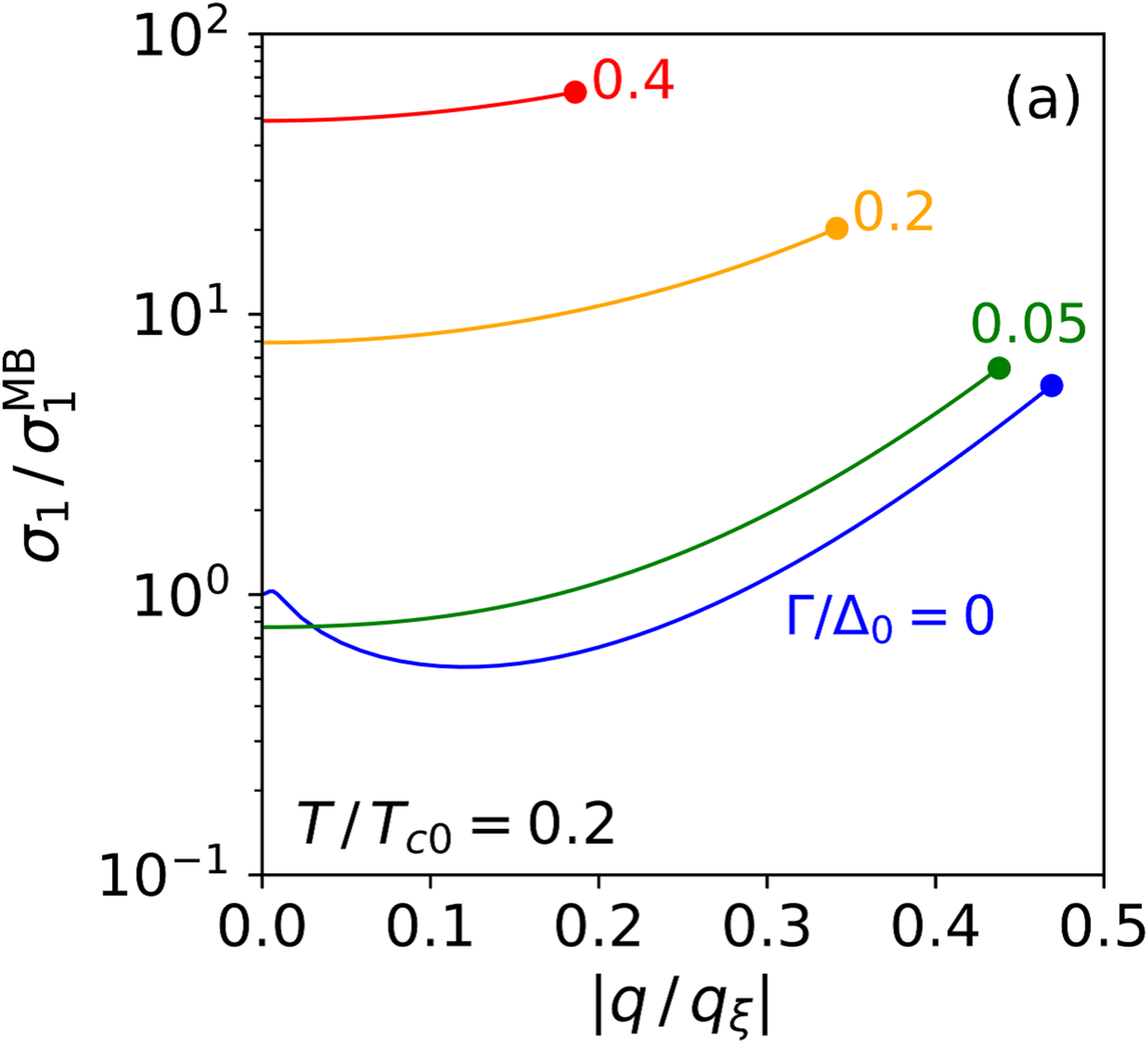}
   \includegraphics[width=0.494\linewidth]{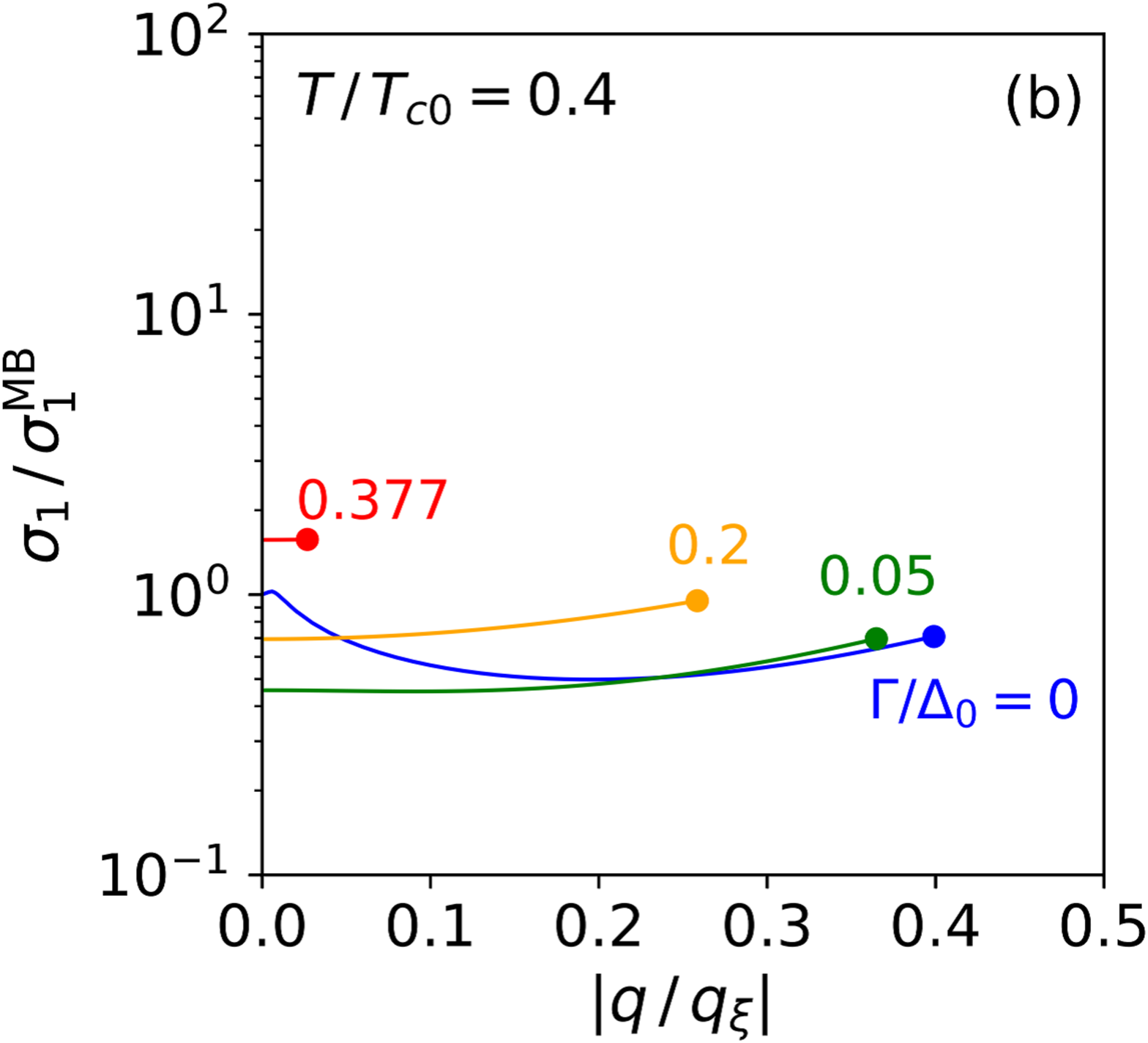}
   \end{center}\vspace{0 cm}
   \caption{
The dissipative conductivity $\sigma_1$ as functions of the superfluid momentum $|q|$ calculated for $\hbar \omega_{\gamma}/\Delta_0 = 0.002$, $\Gamma/\Delta_0=0, 0.05, 0.2, 0.4$ at (a) $T/T_{c0}=0.2$ and (b) $T/T_{c0}=0.4$. 
At each blob, the dc current reaches the depairing current density $j_d(\Gamma, T)$. 
   }\label{fig6}
\end{figure}

\begin{figure}[tb]
   \begin{center}
   \includegraphics[height=0.9\linewidth]{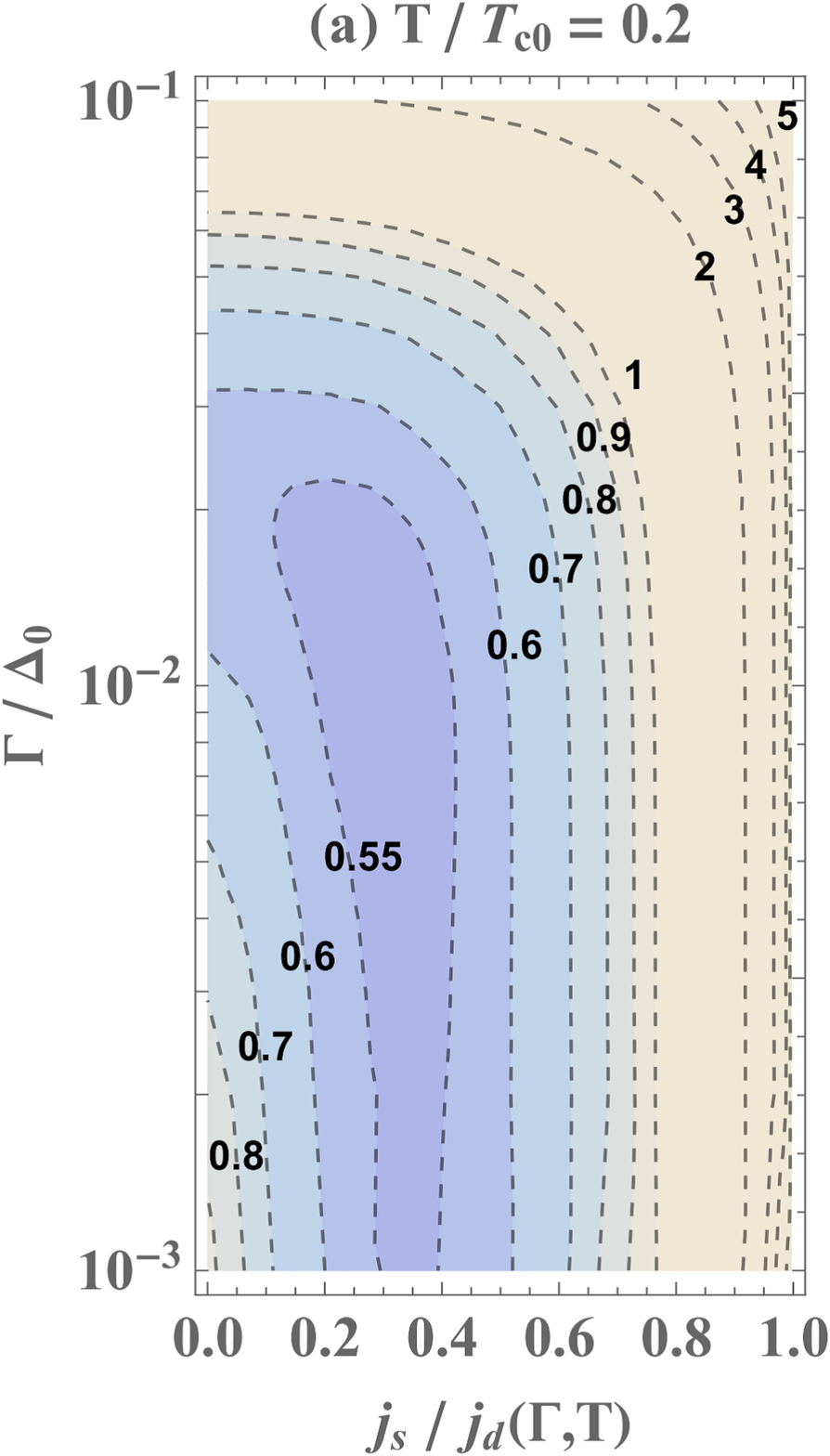}
   \includegraphics[height=0.9\linewidth]{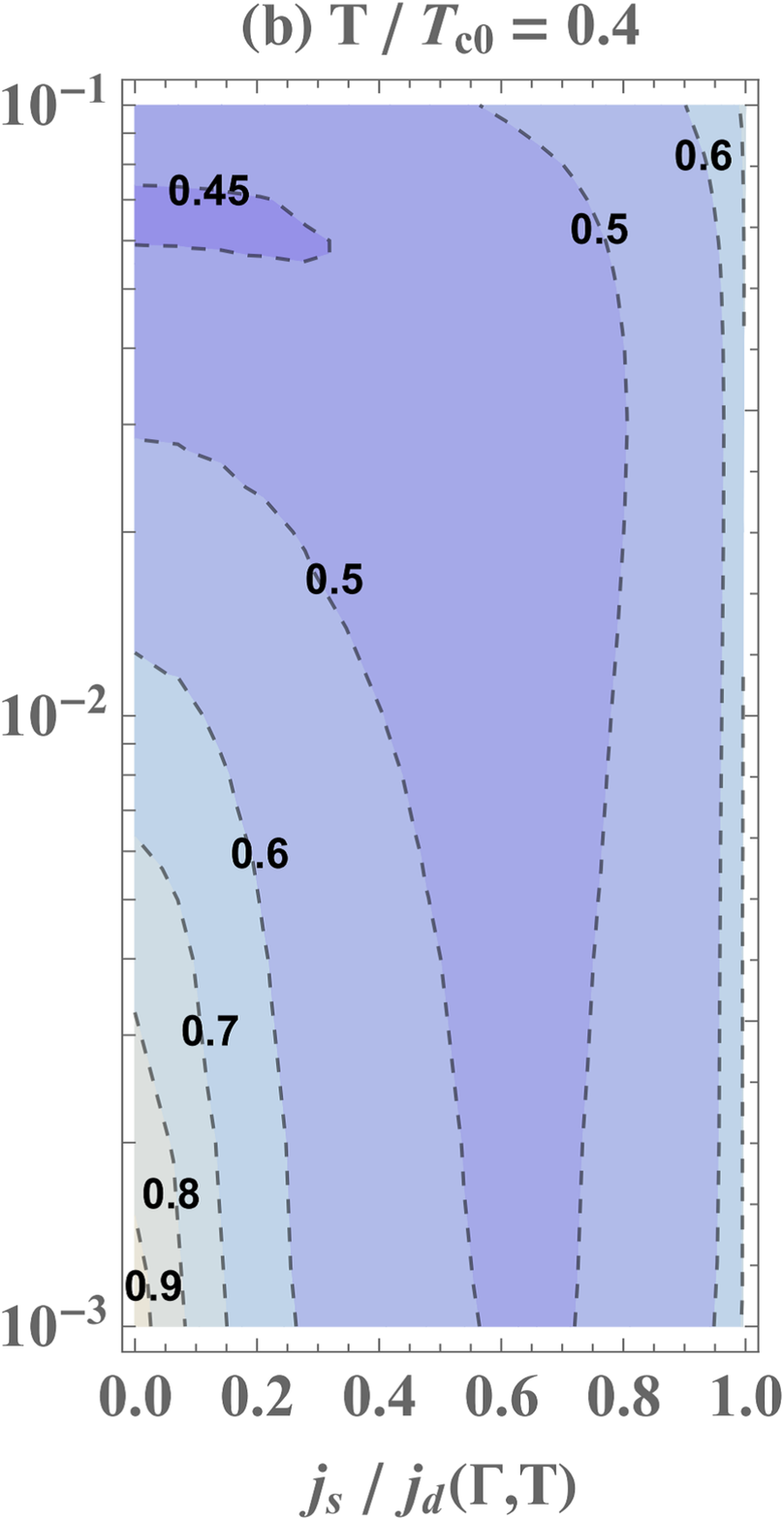}
   \end{center}\vspace{0 cm}
   \caption{
Contour plots of $\sigma_1/\sigma_1^{\rm MB}$ as functions of $j_s$ and $\Gamma$ calculated for $\hbar \omega_{\gamma}/\Delta_0 = 0.002$ at (a) $T/T_{c0}=0.2$ and (b) $T/T_{c0}=0.4$. 
   }\label{fig7}
\end{figure}

The pair-breaking current and an finite $\Gamma$ can strongly affect $\sigma_1$ via the modification of the quasiparticle spectrum. 
Consider the case that the dc current $j_s$ is superposed on the weak time-dependent current with the frequency $\omega_{\gamma}$. 
We assume the amplitude of time dependent current is so tiny that it affects neither the quasiparticle spectrum nor the distribution function. 
The dc bias can be uniform (e.g., nanowires) or has a depth dependence (e.g. SRF cavities). 
In either cases, the local $\sigma_1$ is calculated from Eq.~(\ref{sigma_1}). 
Shown in Fig.~\ref{fig6} are $\sigma_1/\sigma_1^{\rm MB}$ at $\omega_{\gamma} = 0.002$ as functions of the superfluid momentum $|q|$ of the dc current. 
Here $\sigma_1^{\rm MB}=\sigma_1|_{q=0,\Gamma =0} \simeq 0.01\sigma_n$ and $0.6\sigma_n$ for $T/T_{c0}=0.2$ and $0.4$, respectively. 
The blobs represent $\sigma_1$ for the depairing current densities. 
The blue curves represent $\sigma_1$ for the ideal dirty BCS superconductor ($\Gamma=0$)
and exhibit the pronounced minimum~\cite{2014_Gurevich}, 
resulting from the interplay of dc-induced broadening of DOS peaks which reduce $\sigma_1$ and the reduction of spectrum gap which increases $\sigma_1$. 
In the other curves ($\Gamma>0$), the minimum shifts to lower $|q|$ regions. 
This comes from the fact~\cite{2019_Kubo_Gurevich} that 
a finite $\Gamma$ broadens the DOS peaks, 
and the optimum broadening of DOS peaks is achieved by a smaller $|q|$ than for $\Gamma=0$. 
The minimum in $\sigma_1$ disappears when $\Gamma \gtrsim \Gamma_c = T^{3/2} \Delta^{-1/2}$~\cite{2019_Kubo_Gurevich}. 
For $T/T_{c0}=0.2$ and $0.4$, we have $\Gamma_c \sim 0.04$ and $0.1$, respectively. 
Shown in Fig.~\ref{fig7} are the contour plots of $\sigma_1/\sigma_1^{\rm MB}$ as functions of $j_s$ and $\Gamma$. 
In the wide range of parameter regions, 
$\sigma_1$ is smaller than $\sigma_1^{\rm MB}$ by $\sim 50\%$.

\begin{figure}[tb]
   \begin{center}
   \includegraphics[width=0.49\linewidth]{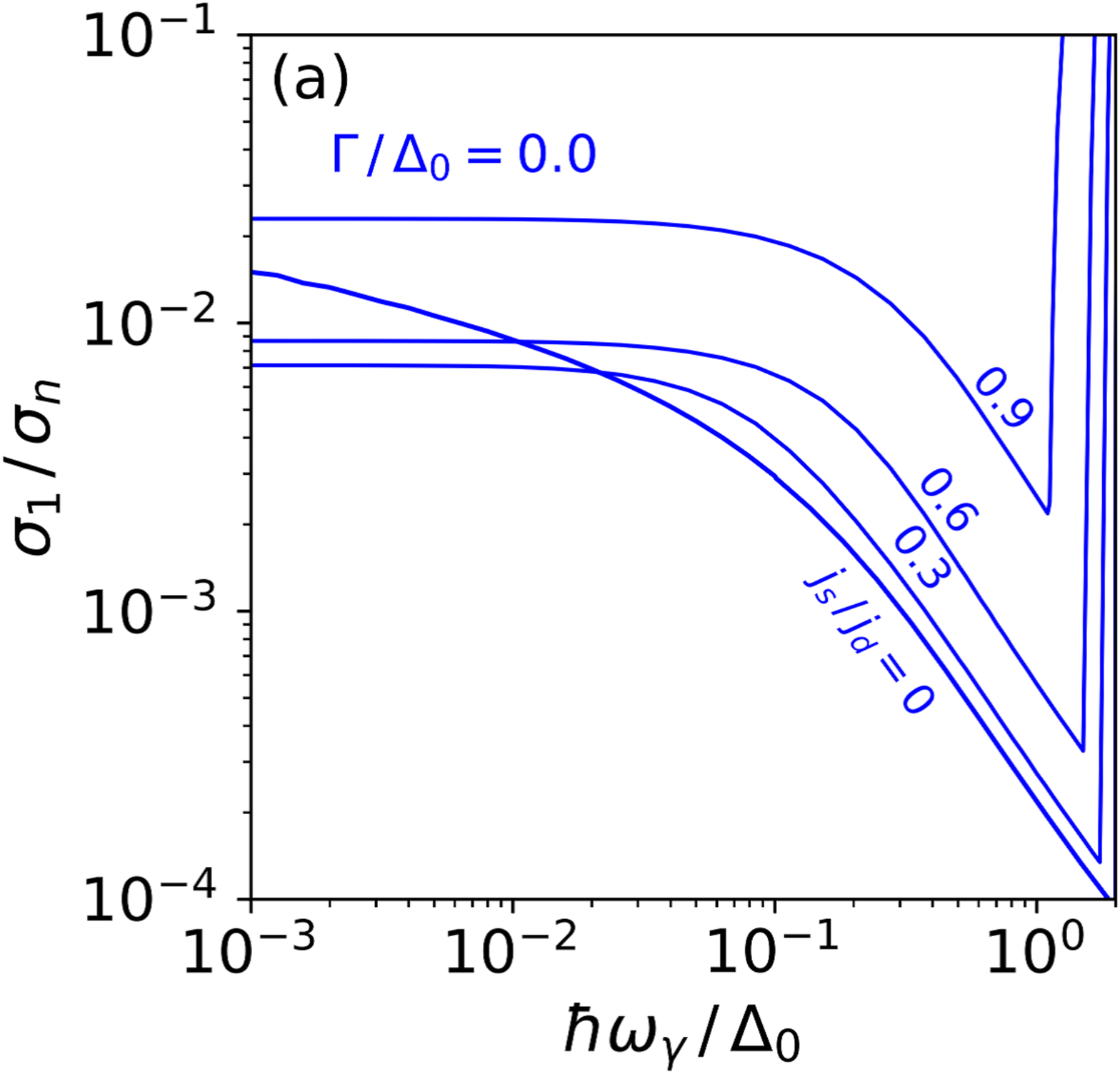}
   \includegraphics[width=0.49\linewidth]{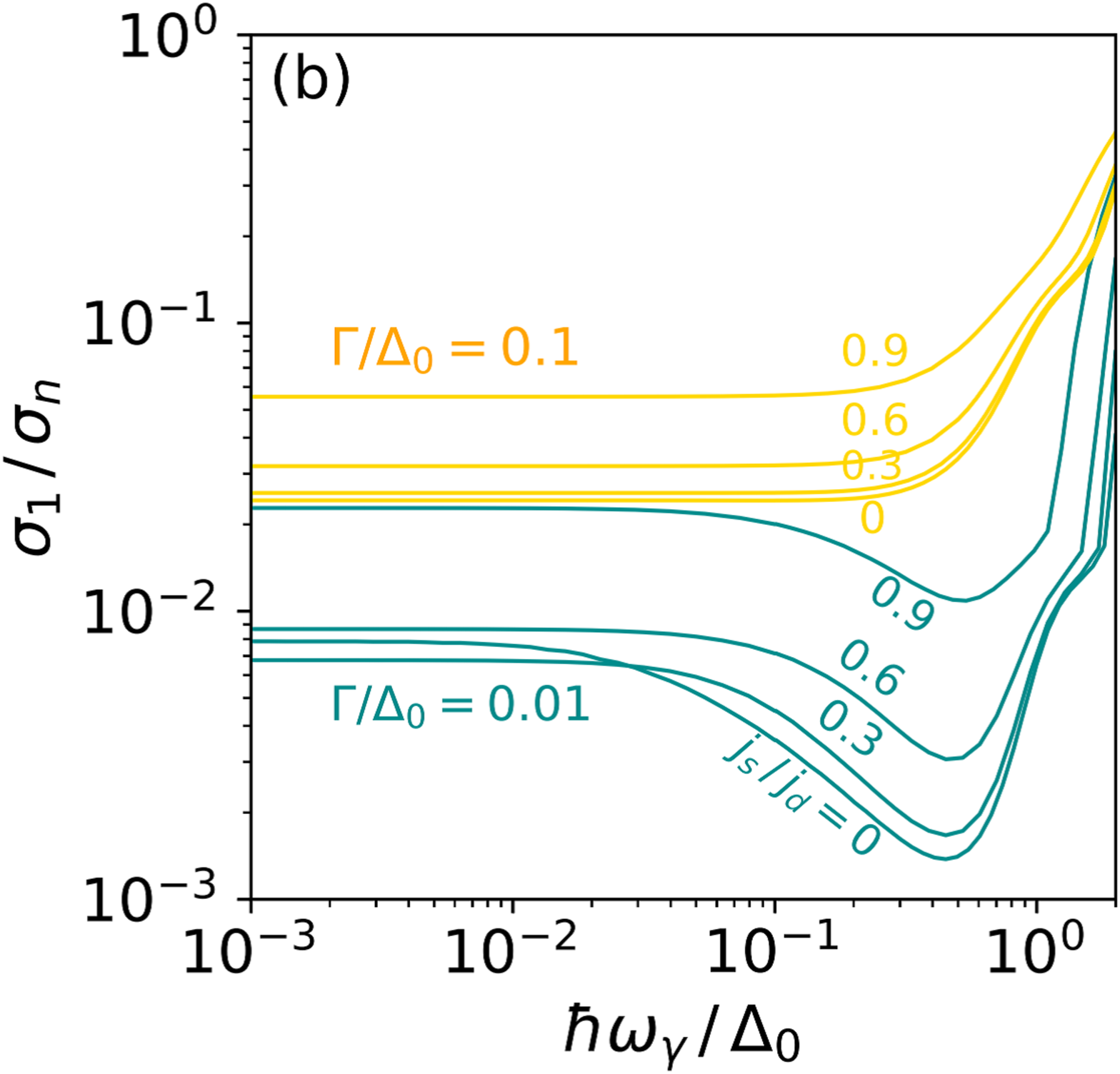}
   \end{center}\vspace{0 cm}
   \caption{
Frequency dependences of $\sigma_1$ at $T/T_{c0}=0.2$ for the dc bias $j_s/j_d = 0, 0.3, 0.6, 0.9$ and 
(a) $\Gamma/\Delta_0 = 0$, (b) 0.01, and 0.1. 
   }\label{fig8}
\end{figure}

For completeness, we discuss the $\omega_{\gamma}$ dependences of $\sigma_1$ under a dc bias $j_s$. 
Shown in Fig.~\ref{fig8} (a) are for $\Gamma=0$ and $j_s \ge 0$. 
When $j_s=0$, we have the well-known logarithmic divergence at $\omega_{\gamma} \to 0$ and the sharp photon-absorption edge at $\hbar \omega_{\gamma} = 2\Delta$. 
For a finite $j_s$, the divergence at $\omega_{\gamma} \to 0$ disappears due to the dc-induced broadening of the DOS peaks. 
As $j_s$ increases, the spectrum gap decreases and the absorption edge shifts to the smaller $\omega_{\gamma}$ regions. 
At $\omega_{\gamma} \ll T$, 
the current can reduce $\sigma_1$ (see also Fig.~\ref{fig6}). 
Shown in Fig.~\ref{fig8} (b) are the $\omega_{\gamma}$ dependences of $\sigma_1$ for $\Gamma>0$ and $j_s \ge 0$. 
In this case, the divergence at $\omega_{\gamma} \to 0$ disappears due to the broadening of the DOS peaks resulting from a finite $\Gamma$ even when $j_s = 0$ (see also Fig.~\ref{fig2}). 
The vague absorption edge appears at around $\Delta$ due to the tail of finite subgap states resulting from $\Gamma>0$, also seen in Fig.~\ref{fig2}. 
As $j_s$ increases, the absorption edge shifts to the smaller $\omega_{\gamma}$ regions. 

\section{Discussion}

We studied in Section~ \ref{zero_current_state} the effects of a finite Dynes $\Gamma$ on various physical quantities in the zero-current state. 
While $T_c$, $\Delta$, $n_s$, $\lambda^{-2}$, and $H_c$ are monotonically decreasing functions of $\Gamma$ (Fig.~\ref{fig1}), 
$\sigma_1$ exhibits a non-monotonic behavior (Fig.~\ref{fig2}). 
A finite $\Gamma$ results in the residual conductivity at lower temperatures, 
but $\sigma_1$ decreases as $\Gamma$ increases due to the broadening of the DOS peaks at $T > (\omega_{\gamma}, \Gamma)$~\cite{2017_Gurevich_SUST, 2017_Gurevich_Kubo}. 
The interplay of the broadening of the DOS peaks, which decreases $\sigma_1$, and the reduction of the spectrum gap, which increases $\sigma_1$, determines the optimum $\Gamma$. 
Then, tuning the quasiparticle spectrum via engineering $\Gamma$ can reduce electromagnetic dissipation in superconducting devices~\cite{2017_Gurevich_SUST, 2017_Gurevich_Kubo}. 
While the physics and materials mechanisms behind $\Gamma$ are not yell understood, 
comparison of tunneling spectroscopy and various materials treatments can give useful information on how to engineer $\Gamma$.

A more convenient control knob for tuning the quasiparticle spectrum is the pair-breaking dc current~\cite{2014_Gurevich}.  
In Sec.~\ref{current_carrying_state}, the effects of the combination of a Dynes $\Gamma$ and a dc bias $j_s$ on the physical quantities are calculated for all $\Gamma$ and all currents up to the depairing current density $j_d$ (Figs.~\ref{fig3}-\ref{fig8}). 
There exists the optimum combination of $\Gamma$ and $j_s$ that minimize $\sigma_1$ (Fig.~\ref{fig7}). 
The minimum value is smaller than that of the ideal dirty zero-current state BCS superconductor by $\sim 50\%$. 
Our results suggest it is possible to minimize dissipation in superconducting devices. 
Once $\Gamma$ for device materials is extracted from tunneling spectroscopy, 
we can reduce $\sigma_1$ by tuning the dc bias along the abscissa of Fig.~\ref{fig7}. 
If it is possible to engineer $\Gamma$ by combining tunneling spectroscopy and various materials processing, 
even more reduction of $\sigma_1$ would be possible by tuning $\Gamma$ along the ordinate of Fig.~\ref{fig7}.

The effect of $\Gamma$ on $\sigma_1$ manifests itself not only in the $j_s$ dependence of $\sigma_1$ but also in the $T$ and the $\omega_{\gamma}$ dependences of $\sigma_1$. 
As shown in Fig.~\ref{fig2} (c) and Fig.~\ref{fig8}, 
the second photon-absorption edge appears at $\omega_{\gamma} \simeq \Delta$, 
which represents the existence of the tail of subgap states. 
As shown in Fig.~\ref{fig2} (b) and  Fig.~\ref{fig6}, 
the height of the coherence peak in $\sigma_1(T)$ is linked to the depth of the minimum in $\sigma_1(j_s)$ through $\Gamma$: 
both are suppressed as $\Gamma$ increases.

We calculated the depairing current density $j_d (\Gamma, T)$ for all $T$ and all $\Gamma$. 
Our results show that $j_d$ is coincident with the Kupriyanov-Lukichev theory~\cite{1980_Kupriyanov} for $\Gamma=0$, but it decreases as $\Gamma$ increases (Fig.~\ref{fig4}).  
So, we can expect that real materials, which usually have $\Gamma>0$, 
exhibit smaller $j_d$ than the prediction by Kupriyanov and Lukichev. 
This is qualitatively consistent with the measurements~\cite{1982_Romijn,2004_Rusanov}, 
but the relation between $j_d$ and $\Gamma$ is still unclear. 
In practice, other mechanisms prevent a precision measurement of $j_d$, 
e.g., current crowding suppresses $\Delta$ and $n_s$ at sharp corners, leading to a smaller $j_d$ than the theoretical values~\cite{2011_Clem}. 
Yet, simultaneous measurement of $j_d$ and $\Gamma$ can lead to a deeper insight into $j_d$ and finding better materials treatment for reducing $\Gamma$ and improving $j_d$.


\begin{acknowledgments}
I would like to express the deepest appreciation to Alex Gurevich for his hospitality during my visit to Old Dominion University. 
This work was supported by Japan Society for the Promotion of Science (JSPS) KAKENHI Grants No. JP17H04839, No. JP17KK0100, and JP19H04395. 
\end{acknowledgments}


\begin{thebibliography}{99}

\bibitem{2017_Padamsee}
H. Padamsee, 
Supercond. Sci. Technol. {\bf 30}, 053003 (2017). 

\bibitem{2017_Gurevich_SUST}
A. Gurevich, 
Supercond. Sci. Technol. {\bf 30}, 034004 (2017). 

\bibitem{2012_Zmuidzinas}
J. Zmuidzinas, 
Annu. Rev. Condens. Matter Phys. {\bf 3}, 169 (2012).

\bibitem{2013_Devoret}
M. H. Devoret and R. J. Schoelkopf, 
Science {\bf 339}, 1169 (2013).

\bibitem{2015_Engel}
A. Engel, J. J. Renema, K. II'in, and A. Semenov,  
Supercond. Sci. Technol. {\bf 28}, 114003 (2015).

\bibitem{2014_Romanenko}
A. Romanenko, A. Grassellino, A. C. Crawford, D. A. Sergatskov, and O. Melnychuk, 
Appl. Phys. Lett. {\bf 105}, 234103 (2014). 

\bibitem{2017_Romanenko}
A. Romanenko and D. I. Schuster
Phys. Rev. Lett. {\bf 119}, 264801 (2017). 

\bibitem{1958_MB}
D. C. Mattis and J. Bardeen, 
Phys. Rev. {\bf 111}, 412 (1958). 

\bibitem{2003_Zasa}
J. Zasadzinski, 
Tunneling spectroscopy of conventional and unconventional superconductors, 
in {\it The Physics of Superconductors}, 
edited by K. H. Bennemann and J. B. Ketterson (Springer, Berlin, 2003), Vol. 1, p. 591.


\bibitem{1978_Dynes}
R. C. Dynes, V. Narayanamurti, and J. P. Garno, Phys. Rev. Lett. {\bf 41}, 1509 (1978).

\bibitem{1984_Dynes}
R. C. Dynes, J. P. Garno, G. B. Hertel, and T. P. Orlando, 
Phys. Rev. Lett. {\bf 53}, 2437 (1984). 

\bibitem{1969_Maki_Parks}
K. Maki, Gapless superconductivity, 
in {\it Superconductivity}, edited by R. D. Parks (Marcel Dekker, Inc., New York, 1969), vol. 2, p. 1035. 

\bibitem{1964_Maki_current}
K. Maki, Prog. Theor. Phys. {\bf 31}, 731 (1964).

\bibitem{1965_Fulde_current}
P. Fulde, Phys. Rev. {\bf 137}, A783 (1965). 

\bibitem{2003_Anthore}
A. Anthore, H. Pothier, and D. Esteve, 
Phys. Rev. Lett. {\bf 90}, 127001 (2003).

\bibitem{1966_Fulde_Maki_mag}
P. Fulde and K. Maki, Phys. Rev. 141, 275 (1966). 

\bibitem{1996_Belzig}
W. Belzig, C. Bruder, and G. Schon, 
Phys. Rev. B {\bf 54}, 9443 (1996).

\bibitem{1999_Belzig}
W. Belzig, F. K. Wilhelm, C. Bruder, G. Schon, and A. D. Zaikin, 
Superlattices Microstruct. {\bf 25}, 1251 (1999).


\bibitem{2014_Gurevich}
A. Gurevich, 
Phys. Rev. Lett. {\bf 113}, 087001 (2014). 

\bibitem{2016_Semenov}
A. V. Semenov, I. A. Devyatov, P. J. de Visser, and T. M. Klapwijk, 
Phys. Rev. Lett. {\bf 117}

\bibitem{2014_Ciovati_Dhakal_Gurevich}
G. Ciovati, P. Dhakal, and A. Gurevich, 
Appl. Phys. Lett. {\bf 104}, 092601 (2014). 

\bibitem{2017_Gurevich_Kubo} 
A. Gurevich and T. Kubo, 
Phys. Rev. B {\bf 96}, 184515 (2017). 

\bibitem{2019_Kubo_Gurevich} 
T. Kubo and A. Gurevich, 
Phys. Rev. B {\bf 100}, 064522 (2019). 

\bibitem{2012_Antoine}
C. Z. Antoine, 
{\it Materials and Surface Aspects in the Development of SRF Niobium Cavities} 
(Institute of Electronic Systems, Warsaw University of Technology, 2012).

\bibitem{2013_Grassellino}
A. Grassellino, A. Romanenko, D. Sergatskov, O. Melnychuk, Y. Trenikhina, A. Crawford, A. Rowe, M. Wong, T. Khabiboulline, and
F. Barkov, Supercond. Sci. Technol. {\bf 26}, 102001 (2013). 

\bibitem{2013_Dhakal}
P. Dhakal, G. Ciovati, G. R. Myneni, K. E. Gray, N. Groll, P. Maheshwari, D. M. McRae, R. Pike, T. Proslier, 
F. Stevie, R. P. Walsh, Q. Yang, and J. Zasadzinzki, 
Phys. Rev. ST Accel. Beams {\bf 16}, 042001 (2013); 
in proceedings of IPAC2012, New Orleans, Louisiana, USA (2012), p. 2426, WEPPC091.

\bibitem{2017_Grassellino}
A. Grassellino, A. Romanenko, Y. Trenikhina, M. Checchin, M. Martinello,
O. S. Melnychuk, S. Chandrasekaran, D. A. Sergatskov, S. Posen,
A. C. Crawford, S. Aderhold, and D. Bice, 
Supercond. Sci. Technol. {\bf 30}, 094004 (2017). 

\bibitem{2017_Maniscalco}
J. T. Maniscalco, D. Gonnella, and M. Liepe, 
J. Appl. Phys. {\bf 121}, 043910 (2017).

\bibitem{2018_Dhakal}
P. Dhakal, S. Chetri, S. Balachandran, P. J. Lee, and G. Ciovati, 
Phys. Rev. Accel. Beams {\bf 21}, 032001 (2018). 

\bibitem{2019_Wenskat}
M. Wenskat, D. Reschke, J. Schaffran, L. Steder, M. Wiencek, D. Bafia, A. Grassellino, O. S. Melnychuk, and A.D. Palczewski, 
in {\it Proceedings of SRF2019, Dresden, Germany}, MOP026 (2019). 

\bibitem{2019_Umemori}
K. Umemori, E. Kako, T. Konomi, S. Michizono, H. Sakai, T. Okada, and J. Tamura, 
in {\it Proceedings of SRF2019, Dresden, Germany}, MOP027 (2019). 


\bibitem{2017_Makita}
J. Makita, J. R. Delayen, A. V. Gurevich, and G. Ciovati, 
in {\it Proceedings of SRF2017, Lanzhou, China}, MOPB035 (2017). 

\bibitem{2019_Maniscalco}
J. T. Maniscalco, T. Gruber, A. T. Holic, and M. Liepe, 
in {\it Proceedings of SRF2019, Dresden, Germany}, TUP051 (2019). 




\bibitem{1981_WattsTobin}
R. J. Watts-Tobin, Y. Krahenbuhl, and L. Kramer, 
J. Low Temp. Phys. {\bf 42}, 459 (1981). 

\bibitem{Kopnin}
N. B. Kopnin, 
{\it Theory of Nonequilibrium Superconductivity.} (Oxford University Press, 2001). 

\bibitem{1980_Kupriyanov}
M. Yu Kupriyanov and V. F. Lukichev, 
Sov. J. Low Temp. Phys. {\bf 6}, 210 (1980). 


\bibitem{2008_Catelani}
G. Catelani and J. P. Sethna, 
Phys. Rev. B {\bf 78}, 224509 (2008).

\bibitem{2012_Lin_Gurevich}
F. Pei-Jen Lin and A. Gurevich, 
Phys. Rev. B {\bf 85}, 054513 (2012). 

\bibitem{2015_Kubo_PTEP}
T. Kubo, 
Prog. Theor. Exp. Phys. {\bf 2015}, 063G01 (2015). 

\bibitem{2017_Kubo_SUST}
T. Kubo, Supercond. Sci. Technol. \textbf{30}, 023001 (2017).

\bibitem{2017_Liarte_SUST}
D. B. Liarte, S. Posen, M. K Transtrum, G. Catelani, M. Liepe, and J. P Sethna, 
Supercond. Sci. Technol. \textbf{30}, 033002 (2017). 

\bibitem{2019_Sauls}
V. Ngampruetikorn and J. A. Sauls, 
Phys. Rev. Research {\bf 1}, 012015 (2019). 

\bibitem{2015_Posen_PRL}
S. Posen, N. Valles, and M. Liepe, 
Phys. Rev. Lett. {\bf 115}, 047001 (2015). 


\bibitem{2006_Gurevich}
A. Gurevich, 
Appl. Phys. Lett. \textbf{88}, 012511 (2006).


\bibitem{2014_Kubo}
T. Kubo,  Y. Iwashita, and T. Saeki, 
Appl. Phys. Lett. \textbf{104}, 032603 (2014). 

\bibitem{2015_Gurevich}
A. Gurevich, 
AIP Adv. \textbf{5}, 017112 (2015).

\bibitem{2016_Tan}
T. Tan, M. A. Wolak, X. X. Xi, T. Tajima, and L. Civale 
Sci. Rep. {\bf 6}, 35879 (2016). 


\bibitem{1970_Usadel}
K. D. Usadel, Phys. Rev. Lett. {\bf 25}, 507 (1970).

\bibitem{2012_Clem_Kogan}
J. R. Clem and V. G. Kogan, 
Phys. Rev. B {\bf 86}, 174521 (2012). 

\bibitem{1967_Nam}
C. B. Nam, 
Phys. Rev. {\bf 156}, 470 (1967). 


\bibitem{2018_Herman}
F. Herman and R. Hlubina, 
Phys. Rev. B {\bf 97}, 014517 (2018). 

\bibitem{2017_Herman}
F. Herman and R. Hlubina, 
Phys. Rev. B {\bf 96}, 014509 (2017). 

\bibitem{1963_Maki_I}
K. Maki, 
Prog. Theor. Phys. {\bf 29}, 10 (1963). 

\bibitem{1963_Maki_II}
K. Maki, 
Prog. Theor. Phys. {\bf 333}, 10 (1963). 



\bibitem{1982_Romijn}
J. Romijn, T. M. Klapwijk, M. J. Renne, and J. E. Mooij, 
Phys. Rev. B {\bf 26}, 3648 (1982). 

\bibitem{2004_Rusanov}
A. Yu Rusanov, M. B. S. Hesselberth, and J. Aarts, 
Phys. Rev. B {\bf 70}, 024510 (2004). 

\bibitem{2011_Clem}
J. R. Clem and K. K. Berggren, 
Phys. Rev. B {\bf 84}, 174510 (2011).

\end{thebibliography}
\end{document}